\documentclass[aps, prx,superscriptaddress, twocolumn,secnumarabic,nobalancelastpage,nofootinbib]{revtex4-1} 
\usepackage{graphicx}
\usepackage{amsmath,amssymb}
\usepackage[utf8]{inputenc}
\usepackage[T1]{fontenc}
\usepackage{lmodern}
\usepackage{natbib}
\usepackage{hyperref}
\usepackage{xcolor}
\hypersetup{colorlinks=true,citecolor={blue},linkcolor={blue},urlcolor={blue}}
\usepackage{units}
\usepackage[english]{babel}
\usepackage[normalem]{ulem}
\usepackage{cancel}
\usepackage{upgreek}
\usepackage{wasysym}
\usepackage{placeins} 
\usepackage{soul} 
\newcommand{\vect}[1]{\mathbf{#1}}
\makeindex

\begin{document}

\definecolor{orange}{RGB}{255, 69, 0}
\definecolor{green}{RGB}{26,148,49}
\setstcolor{red}
\newcommand{\hs}[1]{\textcolor{red}{#1}}
\newcommand{\jdt}[1]{\textcolor{green}{#1}}

\title{Engineering spatial coherence in lattices of polariton condensates}

\date{\today}

\author{J. D. T\"opfer}
\email{J.D.Toepfer@soton.ac.uk}
\affiliation{Skolkovo Institute of Science and Technology, Moscow, Bolshoy Boulevard 30, bld. 1, 121205, Russia}
\affiliation{School of Physics and Astronomy, University of Southampton, Southampton, SO17 1BJ, United Kingdom}

\author{I. Chatzopoulos}
\affiliation{School of Physics and Astronomy, University of Southampton, Southampton, SO17 1BJ, United Kingdom}

\author{H. Sigurdsson}
\affiliation{Skolkovo Institute of Science and Technology, Moscow, Bolshoy Boulevard 30, bld. 1, 121205, Russia}
\affiliation{School of Physics and Astronomy, University of Southampton, Southampton, SO17 1BJ, United Kingdom}

\author{T. Cookson}
\affiliation{School of Physics and Astronomy, University of Southampton, Southampton, SO17 1BJ, United Kingdom}

\author{Y. G. Rubo}
\affiliation{Instituto de Energ\'ias Renovables, Universidad Nacional Aut\'onoma de M\'exico, Temixco, Morelos, 62580, Mexico}

\author{P. G. Lagoudakis}
\email{P.Lagoudakis@skotlech.ru}
\affiliation{Skolkovo Institute of Science and Technology, Moscow, Bolshoy Boulevard 30, bld. 1, 121205, Russia}
\affiliation{School of Physics and Astronomy, University of Southampton, Southampton, SO17 1BJ, United Kingdom}

\renewcommand{\abstractname}{} 
\begin{abstract}
Artificial lattices of coherently coupled macroscopic states are at the heart of applications ranging from solving hard combinatorial optimisation problems to simulating complex many-body physical systems. The size and complexity of the problems scales with the extent of coherence across the lattice. Although the fundamental limit of spatial coherence depends on the nature of the couplings and lattice parameters, it is usually engineering constrains that define the size of the system. Here, we engineer polariton condensate lattices with active control on the spatial arrangement and condensate density that result in near-diffraction limited emission, and spatial coherence that exceeds by nearly two orders of magnitude the size of each individual condensate. We utilise these advancements to unravel the dependence of spatial correlations between polariton condensates on the lattice geometry.
\end{abstract}

\maketitle
\section{Introduction}
Synchronisation and the emergence of coherence between coupled elements are universal concepts arising in nature and technology~\cite{pikovsky2003synchronization}. They dictate collective human behaviour~\cite{strogatz2005crowd}, functioning of neurological systems~\cite{gray1989oscillatory}, as well as phase transitions of quantum systems to macroscopically collective entities at low temperatures~\cite{davis1995bose}. For small-size systems such as two mechanical pendulums coupled through a common support~\cite{bennett2002huygens} or two coupled laser cavities~\cite{soriano2013complex} frequency locking of the two elements depends on the inter-element coupling strength in competition with any inherent dephasing mechanisms. In larger systems consisting of many interacting elements, such as social structures or coupled laser networks the underlying coupling topology (or network architecture) critically influences the dynamics and coherence formed in these systems~\cite{watts_collective_1998, strogatz2001exploring}. Engineering spatial coherence in such large networks is a key element for increasing system performances in power grids~\cite{motter2013spontaneous}, novel computational devices for classification tasks~\cite{romera2018vowel} or laser arrays for creation and control of high power beams~\cite{kao2016phase}.

Lattices of coupled condensates are investigated for the simulation and computation of complex tasks in atomic~\cite{Morsch_RevModPhy2006, struck2011quantum, Bloch_NatPhy2012}, photonic~\cite{dung2017variable} and polaritonic~\cite{amo2016exciton, berloff_realizing_2017, Ballarini_NanoLett2020} platforms. Functionality and computational performance of these systems is ultimately limited by the system's spatial coherence length, i.e. how many condensates can coherently be coupled. From a more fundamental point of view, increasing the coherence allows one to construct a larger many-body system and approach the ideal limit of a homogeneous ``infinite'' order parameter, which is essential for the study of phase transitions in interacting bosonic systems such as the Berezinskii-Kosterlitz-Thouless transition in two-dimensions (2D)~\cite{Hadzibabic_Nature2006}. In the case of exciton-polariton condensates in semiconductor microcavites~\cite{Kasprzak_Nature2006}, 
measurements were initially limited to small condensate sizes~\cite{Deng_PRL2007, Krizhanovskii_PRB2009, Manni_PRL2011, Roumpos_PNAS2012} but advancements in sample fabrication and experimental techniques now allow studies on their coherence properties as extended objects far away from the excitation area~\cite{Wertz_NatPhys2010, Caputo_NatMat2018}, in optical traps~\cite{Ohadi_PRB2018,askitopoulos2019giant}, and photonic microstructures~\cite{Baboux_Optica2018}. The emergence of correlations beyond the spatial extension of the laser excitation area has underpinned the creation of networks and lattices of polariton condensates.

Here, we engineer control over the particle density and position of each polariton condensate across a polariton lattice under non-resonant optical excitation. We overcome the challenge of disorder induced localisation and dephasing by employing a feedback scheme for each individual laser spot, balancing the condensate density and allowing us to accurately study the decay of coherence across different coupling topologies. The result is a homogeneous macroscopic, high energy, condensate lattice with coherence length exceeding multiple lattice cells and near-diffraction limited emission. Generating such large condensate lattices by using multiple excitation sources allows us to go beyond the standard single-excitation source limit~\cite{Wertz_NatPhys2010, Caputo_NatMat2018} overcoming beam size limitations, beam profile inhomogeneities, and condensate fragmentation~\cite{Kasprzak_Nature2006}. We observe that the connectivity of the lattice significantly enhances the system's coherence length due to increased coherent coupling between adjacent condensates.

\section{Results}
We optically generate lattices of coupled polariton condensates using non-resonant, pulsed and tightly-focused laser excitation spots for each condensate node (see Methods for details). Polaritons generated at each condensate node convert inherited potential energy into kinetic energy resulting in a radially (ballistically) expanding polariton fluid from an antenna like source~\cite{Richard_PRL2005, topfer2020time}. When two or more expanding condensates are brought together, interference effects are revealed implying phase synchronisation between the condensation centres and the emergence of a macroscopic order parameter~\cite{tosi_sculpting_2012, Cristofolini_PRL2013, Ohadi_PRX2016, berloff_realizing_2017, topfer2020time,Alyatkin_PRL2020}.

The spatial coherence in a network (or lattice) of $N$ polariton condensates generated under pulsed excitation is described by the integrated complex coherence factor (see Appendix~\ref{appendix_Coherence})
\label{eq.integratedcomplexcoherencefactor}
\begin{equation}
\tilde{\mu}_{ij} =\frac{  \int \Gamma_{ij}(t) \mathrm{d}t }{ \sqrt{  \int  \Gamma_{ii}(t)  \mathrm{d}t   \int  \Gamma_{jj}(t)  \mathrm{d}t } }, \quad i,j = 1,...,N.
\end{equation}
Here, the correlation function $\Gamma_{ij}(t) = \left\langle \psi_i(t)^*  \psi_j(t) \right\rangle $ denotes the mutual intensity of each pair of condensates averaged over many realisations (pulses) of the system, and $\psi_i(t)$ is the complex-valued amplitude of the $i \mathrm{th}$ condensate. While the modulus of the complex coherence factor $|\tilde{\mu}_{ij}| \leq 1$ is a normalised measure for the coherence between two condensate nodes its argument $\tilde{\theta}_{ij} = \arg(\tilde{\mu}_{ij})$ represents their average phase-difference.
\subsection{Condensate density stabilisation}
\begin{figure}[!t]
	\center
	\includegraphics[]{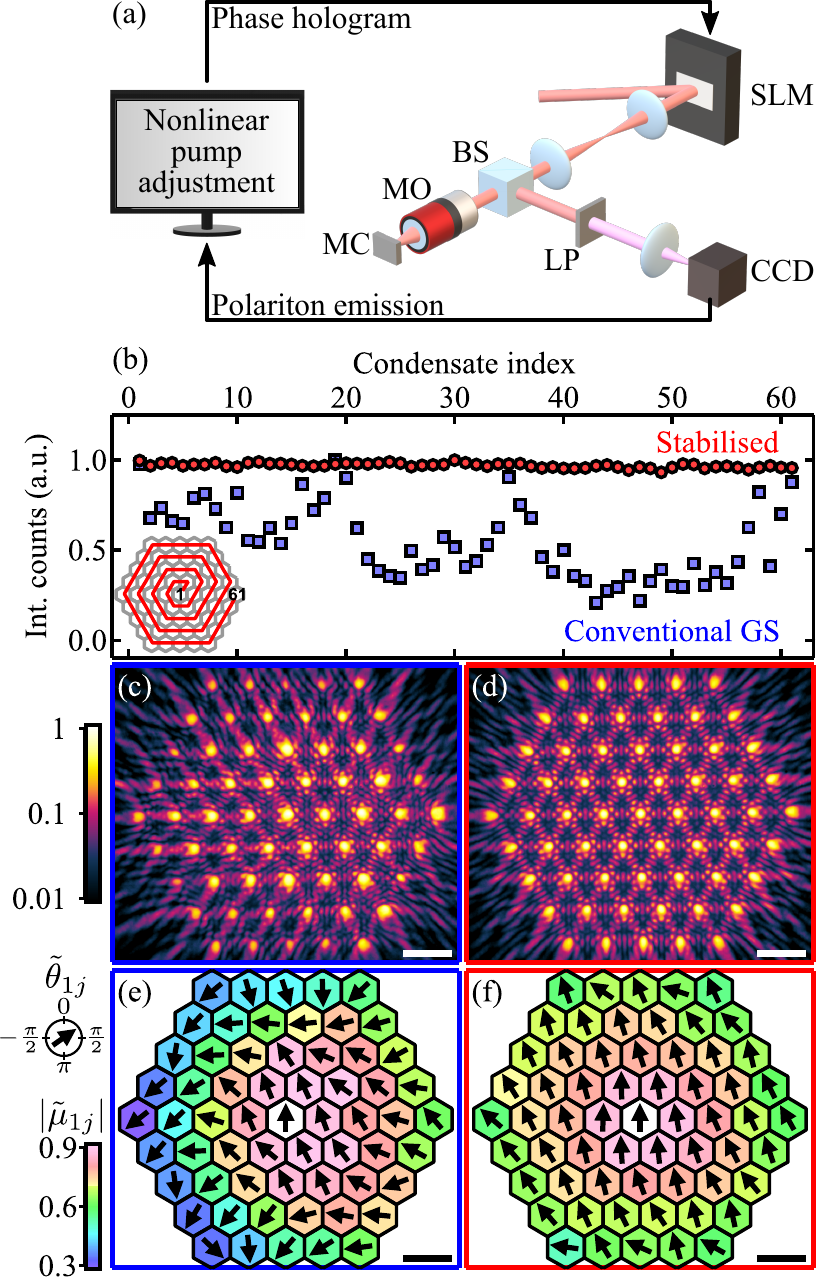}
	\caption{Closed-loop density stabilisation in polariton lattices. (a) Schematic of the feedback loop that iteratively analyses the emission and adjusts the pump profile to equalise the emission intensities of all condensate nodes. (b) Integrated emission of 61 condensate nodes at condensation threshold ($P=P_\mathrm{thr}$) in a triangular lattice configuration (lattice constant $a=14.9\;\mathrm{\upmu m}$) without (blue squares) and with (red circles) density stabilisation. Inset depicts the condensate number indexing. (c,d) Recorded real-space photoluminescence and (e,f) measured complex coherence factor $\tilde{\mu}_{1j}$ between the central condensate node $1$ and each other condensate node $j$ without and with density stabilisation, respectively. False colourscale and pseudo-spins (black arrows) depict magnitude $|\tilde{\mu}_{1j}|$ and phase $\tilde{\theta}_{1j}$. Data shown in (c-f) are extracted at a total pump power $P=1.2P_{\mathrm{thr}}$. Scale bars in (c-f) correspond to $20\;\mathrm{\upmu m}$. Abbreviations in (a): SLM - spatial light modulator, BS - beam splitter, MO - microscope objective lens, MC - microcavity, LP - longpass filter, CCD - Charge-coupled device.}
	\label{Fig1}
\end{figure}

We utilise a reflective liquid-crystal phase-only spatial light modulator (SLM) to modulate the Gaussian excitation pump beam in the Fourier plane of the optical excitation system to generate desired excitation pump spot geometries at the focal plane of the microscope objective lens (see schematic in Fig.~\ref{Fig1}(a)).
Phase holograms are calculated using a modified version of the well-known Gerchberg-Saxton (GS) algorithm~\cite{gerchberg1972practical}, in which the measured polariton photoluminescence (PL) of the condensate lattice feeds back into the closed-loop sequence (see Appendix~\ref{appendix_FeedbackScheme} for a full description of the method). Iterative nonlinear adjustment of the excitation pump profile allows us to create macroscopic lattices of $>100$ condensates with homogeneous condensate node density ($\leq 1\%$ relative standard deviation (RSD)) across the whole lattice.

In Fig.~\ref{Fig1}(b) we compare the measured emission of all condensate nodes in a triangular lattice of 61 elements and lattice constant $a=14.9\;\mathrm{\upmu m}$ when using $100$ iterations of the conventional GS algorithm (blue squares) and the presented condensate density stabilisation method (red circles), respectively. Unavoidable effects such as sample disorder, finite accuracy of the GS algorithm, optical aberrations and missing translational invariance due the finite size of the lattice~\cite{kalinin2018networks} all contribute to a broad distribution of effective gain for each condensate node and, thus, to a large spread $\approx 37\%$ (RSD) in the distribution of condensate emission powers for the case of no condensate density stabilisation. Active stabilisation of the lattice using the described closed-loop sequence allows us to compensate for these detrimental elements and yield a reduced spread $\approx 1\%$ (RSD), which is limited by experimental noise in the system.

The near-field (or real space) polariton photoluminescence (PL) measured at pump power $P=1.2P_\mathrm{thr}$, where $P_\mathrm{thr}$ is the system's condensation threshold pump level, is shown in Figs.~\ref{Fig1}(c) and (d) for both excitation schemes. The presence of interference fringes with large visibility in-between ballistically expanding condensates demonstrates mutual coherence between nearest neighbour condensates. However, the lack of homogeneity in the spatial distribution of interference fringes in Fig.~\ref{Fig1}(c) for the case of no density stabilisation indicates a broad distribution of relative phase-difference $\theta_{ij}$ between nodes. This is further confirmed by detailed measurements of the integrated complex coherence factor $\tilde{\mu}_{ij}$ (see Appendix~\ref{appendix.FarfieldMeasurementTechnique} and \ref{appendix.MeasurementOfCoherence} for methods). In Figs.~\ref{Fig1}(e) and (f) we plot magnitude $|\tilde{\mu}_{1j}|$ and phase $\tilde{\theta}_{1j}$ between the central condensate node $1$ and each other condensate node $j$ using false-colour and pseudo-spins (black arrows). We find an enhanced and isotropic spatial decay of coherence $|\tilde{\mu}_{1j}|$ and larger homogeneity in relative phase-differences $\tilde{\theta}_{1j}$ for the density stabilised polariton lattice. The noticeable increase of phase-differences $\tilde{\theta}_{1j}$ - or analogous rotation of pseudo-spins - towards the edges of the lattice is an expected finite-size effect due to a flux of particles escaping the system~\cite{kalinin2018networks}. In the ideal scenario of an infinite triangular lattice, with homogeneous condensate occupation numbers, one retrieves a homogeneous distribution of phase differences due to the system's translational invariance (see Appendix~\ref{subsection_InfiniteLatticeSize}).

\begin{figure}[!t]
	\center
	\includegraphics[]{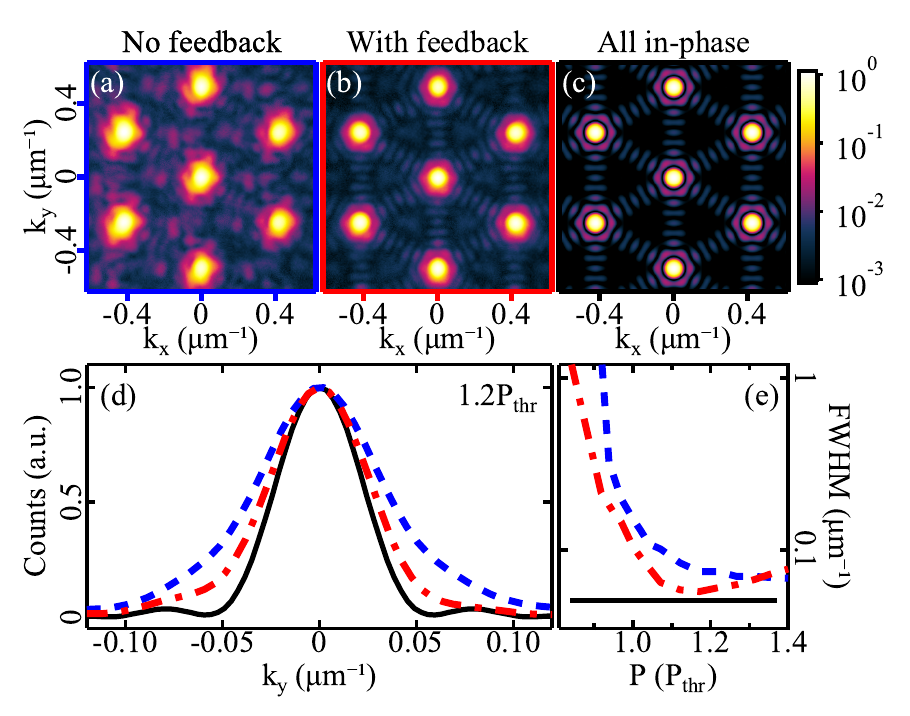}
	\caption{Emission characteristics of a triangular polariton lattice in reciprocal space. First Brillouin zones of the far-field emission for a triangular polariton lattice of 61 condensates excited (a) without feedback, (b) with feedback and for (c) a calculated ideal system of 61 coherent phase-synchronised point sources. (d) Extracted profiles of the central Bragg peak along $k_x=0$ for the far-field emission patterns shown in (a-c). (e) Extracted FWHM of the central Bragg-peak as a function of total excitation pump power. Data shown in (a,b,d) are recorded at pump power $P=1.2P_\mathrm{thr}$.}
	\label{Fig2}
\end{figure}

The polariton condensate lattice is further probed by far-field (or reciprocal space) measurements in analogy to time-of-flight experiments in cold atom systems~\cite{Hadzibabic_Nature2006} (see Appendix~\ref{appendix.FarfieldMeasurementTechnique} for methods). A zoom into the first Brillouin zones of the recorded far-field emission pattern is shown in Figs.~\ref{Fig2}(a) and (b) for the two lattice realisations with and without node density stabilisation pumped at $P=1.2P_\mathrm{thr}$. For comparison, we illustrate in Fig.~\ref{Fig2}(c) the calculated far-field emission of 61 superimposed fully-coherent point-sources (wavefunctions) in triangular arrangement which shows good agreement with the emission pattern of the stabilised condensate lattice. We compare the width of the central far-field emission peak at $k=0$ for the three cases shown in Figs.~\ref{Fig2}(a-c) by plotting the corresponding extracted intensity profiles along $k_x=0$ in Fig.~\ref{Fig2}(d). The lobe of the calculated fully-coherent system (black line) represents the diffraction limited interference peak. We extract the central lobe's full-width-at-half-maximum (FWHM) for varying pump power $P$ as shown in Fig.~\ref{Fig2}(e) and find near-diffraction limited far-field emission for the density stabilised lattice (red dash-dotted line), i.e. a minimum width at $P=1.17P_\mathrm{thr}$ which is only $\approx 13\%$ larger than the diffraction limit of a fully-coherent system (black horizontal line). Without active condensate density stabilisation (blue dashed line) the peak width increases to $\approx 47\%$ at the same pump power. The observed broadening of the far-field emission peak in the physical system is a result of both reduced coherence $|\tilde{\mu}_{ij}|$ and non-homogeneous phase distribution $\tilde{\mu}_{ij}$. The system's pump-power dependencies are summarised in Appendix~\ref{appendix_PulsedPowerDependence}.

\subsection{Coherence vs. dimensionality}
\label{section_dimensionality}
In this section we begin by investigating the coherence between two ballistic polariton condensates. Previous studies of this system have revealed periodically alternating synchronisation patterns of in-phase ($\tilde{\theta}_{12}=0$) and anti-phase ($\tilde{\theta}_{12}=\pi$) states with increasing condensate separation distance $d_{12}$~\cite{Ohadi_PRX2016,topfer2020time}. While there exist separation distances at which the emission of the coupled condensate system is not single-mode but exhibits two or more modes of both even and odd parity states~\cite{Ohadi_PRX2016, topfer2020time}, in the following, we focus on single-mode realisations only. Under this condition the integrated complex coherence factor $\tilde{\mu}_{12}$ is a measure for the system's average coherence properties. In Appendix~\ref{appendix_TimeResolvedEmission} we detail on the time-resolved coherence build-up $\mu_{12}(t)$ of a polariton dyad.
\begin{figure}[!t]
	\center
	\includegraphics[]{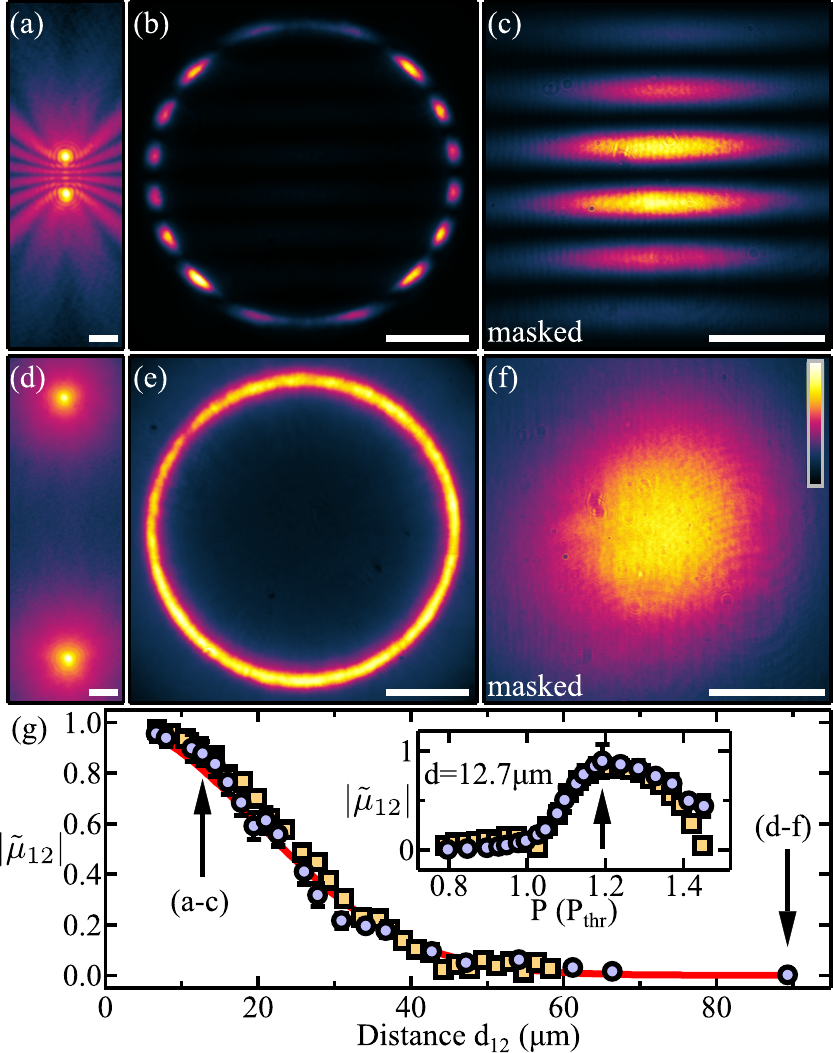}
	\caption{Ballistic expansion, coupling and interference of two polariton condensates. Recorded (a,d) real-space and (b,c) far-field PL of two condensates with separation distances $d_{12} = 12.7\;\mathrm{\upmu m}$ and $d_{12} = 89.3\;\mathrm{\upmu m}$, respectively. Corresponding far-field interference patterns after masking of the emission in real space to block all emission outside the $2\;\mathrm{\upmu m}$ FWHM of each condensate nodes are shown in (c) and (f). (g) Distance dependence of the extracted  integrated complex coherence factor $|\tilde{\mu}_{12}|$, while keeping the excitation pump power constant at $P = 1.2 P_{\mathrm{thr}}$, where $P_{\mathrm{thr}}$ is the measured threshold pump power at a separation distance of $d_{12} = 12.7\;\mathrm{\upmu m}$. Red curve is a Gaussian fit (Eq.~\eqref{Eq.GaussianDecay}) to the experimental data points. Inset shows the mutual coherence obtained by numerical simulations. False colour scale in (f) applies to (b,c,e,f) in linear scale and to (a,d) in logarithmic scale saturated below $10^{-4}$ of the maximum count rate. Scale bars in (a,d) and (b,c,e,f) 	correspond to $10\;\mathrm{\upmu m}$ and $1\;\mathrm{\upmu m}^{-1}$, respectively.}
	\label{Fig3}
\end{figure}

Recorded near-field and far-field PL of two condensates separated by $d_{12}=12.7\;\mathrm{\upmu m}$ and pumped equally at $P=1.2P_{\mathrm{thr}}$  are shown in Figs.~\ref{Fig3}(a) and (b). The interference patterns in both emission images reveal anti-phase synchronisation between the two condensates ($\tilde{\theta}_{12}=\pi$) due to the even number of interference fringes. Far-field emission consists predominantly of PL at large in-plane wavevector $k=1.8\;\mathrm{\upmu m}^{-1}$ demonstrating the ballistic expansion of both condensates due to the strong repulsive interactions between polaritons and the pump-induced exciton reservoirs localised at each condensate center~\cite{Wouters_2008PRB}. In Fig.~\ref{Fig3}(c) we show the recorded far-field interference pattern when spatially filtering the central $2\;\mathrm{\upmu m}$ FWHM of each condensate. Analysis of this double hole interference pattern reveals a large coherence factor $|\tilde{\mu}_{12}| = 0.9 $. Increasing the separation distance $d_{12}$ while keeping the the pump power $P$ constant leads to a decay of coherence $|\tilde{\mu}_{12}|$ between the two condensate centres such that at $d_{12} = 89.3\;\mathrm{\upmu m}$ no interference pattern can be found in near-field an far-field emission as shown in Figs.~\ref{Fig3}(d-f).

The dependency of the measured coherence $|\tilde{\mu}_{12}|$ on the pump spot separation distance $d_{12}$ is shown in Fig.~\ref{Fig3}(g) as blue circles and is fitted with a Gaussian decay
\begin{equation} \label{Eq.GaussianDecay}
|\tilde{\mu}_{12}(d_{12})| =  \exp{\left\{-\frac{\pi}{4} \left( \frac{d_{12}}{L_c} \right)^2 \right\}}.
\end{equation}
The fit parameter $L_c = 24.9 \pm 0.4\;\mathrm{\upmu m}$ represents a measure for the length over which synchronisation of the two-condensate system is possible, and we denote it as the effective coherence length,
\begin{equation*}
L_c =  \int_0^\infty |\tilde{\mu}_{12}(x)| \mathrm{d}x.
\end{equation*}
We reproduce the experimentally measured decay of mutual coherence between two coupled polariton condensates through 2D numerical simulation using the generalised stochastic Gross-Pitaevskii equation (GPE) shown in Fig.~\ref{Fig3}(g) as yellow squares (see Methods). The dependency of coherence $|\tilde{\mu}_{12}|$ as a function of pump power $P$  between two coupled condensates with separation distance $d_{12}=12.7\;\mathrm{\upmu m}$ is depicted in the inset of Fig.~\ref{Fig3}(g) for both experiment (blue circles) and numerical simulation (yellow squares). An increase of coherence between the two condensate nodes arising at condensation threshold is followed by a drop of $|\tilde{\mu}_{12}|$ for larger pump power $P > 1.2 P_{\mathrm{thr}}$. This decrease in $|\tilde{\mu}_{12}|$ is largely attributed to the transition of the system into multi-mode operation~\cite{Ohadi_PRX2016} yielding reduced visibility in time-integrated measurements. We point out that the system realisations which are shown in the distance-dependence in Fig.~\ref{Fig3}(g) exhibit single-mode emission only. The decrease in coherence with increasing separation distance is caused by the spatial decay of the wavefunction of ballistically expanding polariton condensates~\cite{topfer2020time}, i.e. a reduced coupling strength between the two condensate nodes. In Appendix~\ref{appendix_cw} we compare our results to a dyad under continuous wave excitation.

\begin{figure*}[!t]
	\center
	\includegraphics[]{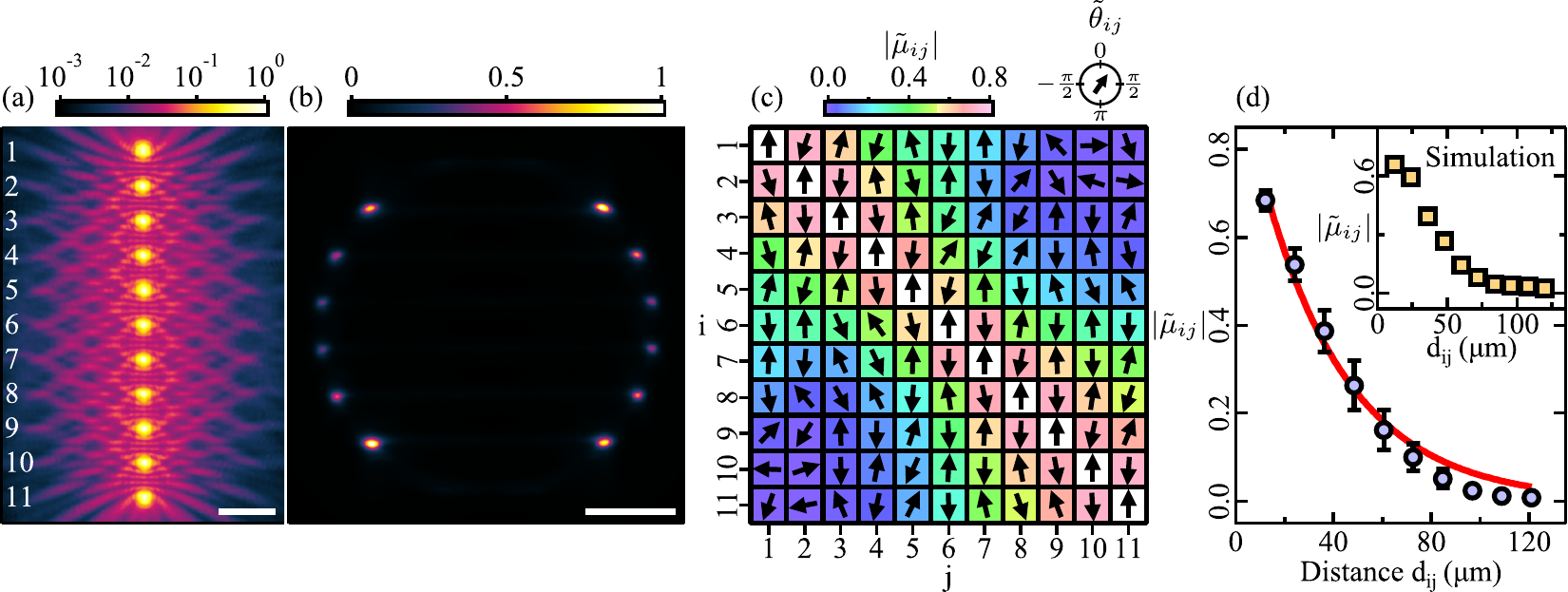}
	\caption{Finite 1D lattice of ballistically coupled polariton condensates. Recorded (a) near-field and (b) far-field PL for 11 condensates arranged in a chain with lattice constant $a=12.1\;\mathrm{\upmu m}$. (c) Correlation matrix illustrating measured mutual coherence factor $|\tilde{\mu}_{ij}|$ and phase factor $\tilde{\theta}_{ij} = \arg{(\tilde{\mu}_{\ij})}$ for each pair of condensates $i \neq j$. Condensate indexing is shown in (a). (d) Spatial decay of the averaged mutual coherence factors $|\tilde{\mu}_{ij}|$ with GPE simulation shown in inset. Error bars represent the standard deviation of the set of coherence factor values for each separation distance $d_{ij}$. Red curve is an exponential fit to the experimental data points. The pump power is kept at $P=1.2 P_{\mathrm{thr}}$ for all data shown in (a-d). Scale bars in (a) and (b) correspond to $20\;\mathrm{\upmu m}$ and $1\;\mathrm{\upmu m}^{-1}$, respectively.}
	\label{Fig4}
\end{figure*}

Next, we increase the number of condensates by investigating a linear chain of 11 equally spaced condensates with nearest-neighbours (NN) distance $a=12.1\;\mathrm{\upmu m}$. The resultant interference patterns in both real and reciprocal space shown in Figs.~\ref{Fig4}(a) and (b) indicate anti-phase synchronisation, i.e. phase differences  between any pair $\{i,j\}$ of condensates according to
\begin{equation} \label{Eq.Phasedifferences_Chain}
\tilde{\theta}_{ij} = \pi \cdot (i-j) \mod\pi,
\end{equation}
with condensate indexing shown in Fig.~\ref{Fig4}(a). Such order can be said to be ``antiferromagnetic''.

We measure and analyse the far-field interference between each pair of condensates $i \neq j$. Extracted magnitude $|\tilde{\mu}_{ij}|$ and phase $\tilde{\theta}_{ij}$ of the integrated complex coherence factor are illustrated in matrix form in Fig.~\ref{Fig4}(c) with row and column indices $i,j$ denoting the pair of condensates, where we make use of the hermiticity of the correlation matrix $\tilde{\mu}_{ij} = \tilde{\mu}^{*}_{ji}$. In agreement with Eq.~\eqref{Eq.Phasedifferences_Chain} we confirm anti-phase synchronisation between NNs (antiparallel pseudo-spins). For large condensate pair distances, $|i-j| \gg 1$, the coherence factor $\tilde{\mu}_{ij}$ decays in magnitude and its phase deviates from Eq.~\eqref{Eq.Phasedifferences_Chain} indicating loss of long-range antiferromagnetic order.
In Fig.~\ref{Fig4}(d) we show the decay of coherence $|\tilde{\mu}_{ij}|$ in the chain as a function of condensate spacing $d_{ij}$ (i.e., lattice constant $a$ is fixed). The data are fitted with a single exponential decay (red curve)
\begin{equation} \label{Eq.ExponentialDecay}
|\tilde{\mu}_{ij}| =  \exp{ \left\{ - d_{ij} / L_c \right\} },
\end{equation}
which yields a coherence length $L_c=35\pm 1\;\mathrm{\upmu m}$. 

\begin{figure*}[!t]
	\center
	\includegraphics[]{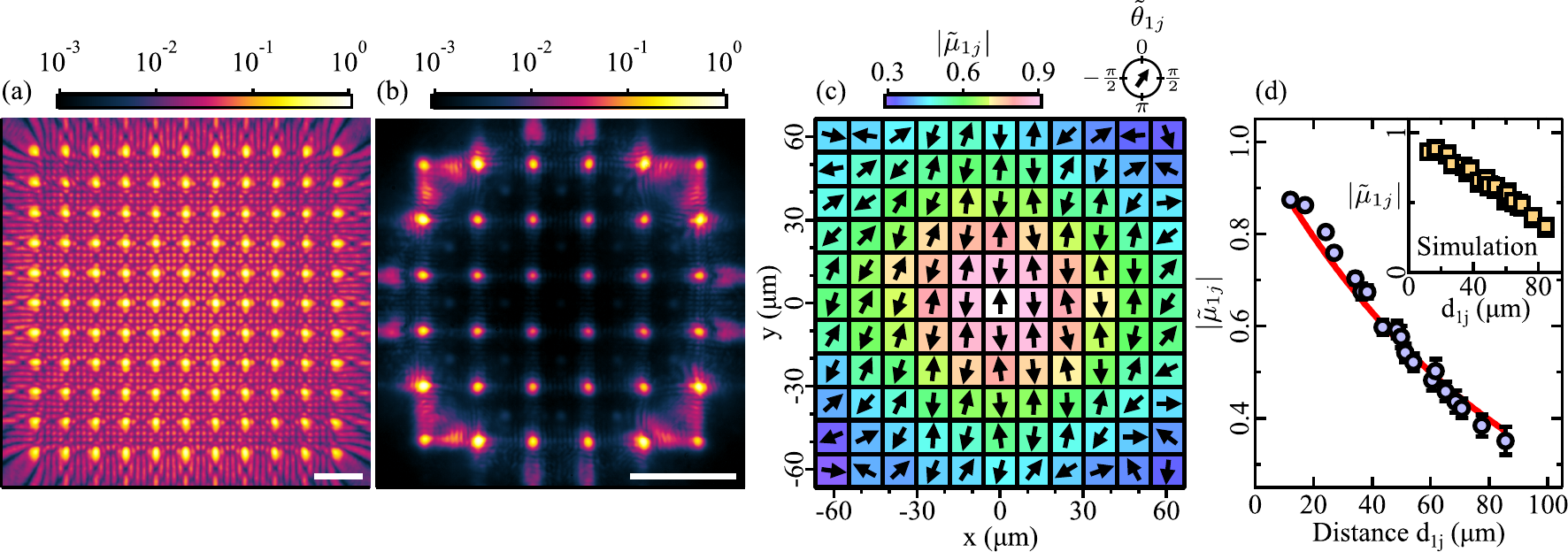}
	\caption{Finite two-dimensional square lattice of ballistically coupled polariton condensates. Recorded (a) near-field and (b) far-field PL for 121 condensates arranged in a square lattice with lattice constant $a=12.1\;\mathrm{\upmu m}$. Spatial correlation map showing mutual coherence factor $|\tilde{\mu}_{1j}|$ and phase factor $\tilde{\theta}_{1j} = \arg{(\tilde{\mu}_{1j})}$ between the most central condensate (index 1) and each other condensate (index $j$). (d) Spatial decay of the averaged mutual coherence factors $|\tilde{\mu}_{1j}|$ with GPE simulation shown in inset. Error bars represent the standard deviation of the set of coherence factor values for each separation distance $d_{ij}$. Red curve is an exponential fit to the experimental data points. The pump power is kept at $P=1.2 P_{\mathrm{thr}}$ for all data shown in (a-d). Scale bars in (a) and (b) correspond to $20\;\mathrm{\upmu m}$ and $1\;\mathrm{\upmu m}^{-1}$, respectively.  }
	\label{Fig5}
\end{figure*}

In a next step, we increase the system size to a 2D square lattice comprising 121 condensates with lattice constant $a = 12.1\;\mathrm{\upmu m}$. The interference patterns in near-field and far-field PL which are shown in Figs.~\ref{Fig5}(a) and (b) reveal anti-phase synchronisation in analogy to the 1D system in Fig.~\ref{Fig4}. We measure $\tilde{\mu}_{1j}$ between the central condensate (index $1$) and all other condensate nodes $j=2,...,121$ plotted in Fig.~\ref{Fig5}(c). The observed antiferromagnetic order ($\tilde{\mu}_{1j}\approx \pm 1)$ reduces with distance from the centre seen from the pseudo-spin rotation towards the edges in Fig.~\ref{Fig5}(c) due to flux of particles escaping the lattice. However, coherence $|\tilde{\mu}_{1j}|$ between the most central condensate and any other condensate of the lattice does not drop below $0.3$ for separation distances as large as $d_{1j} = 86\;\mathrm{\upmu m}$ towards the corners of the lattice. The spatial decay of coherence $|\tilde{\mu}_{1j}|$ is shown in Fig.~\ref{Fig5}(d) and fitted with an exponential function (Eq.\eqref{Eq.ExponentialDecay}, red curve) with coherence length $L_c=87 \pm 1\;\mathrm{\upmu m}$.

\begin{figure*}[!t]
	\center
	\includegraphics[]{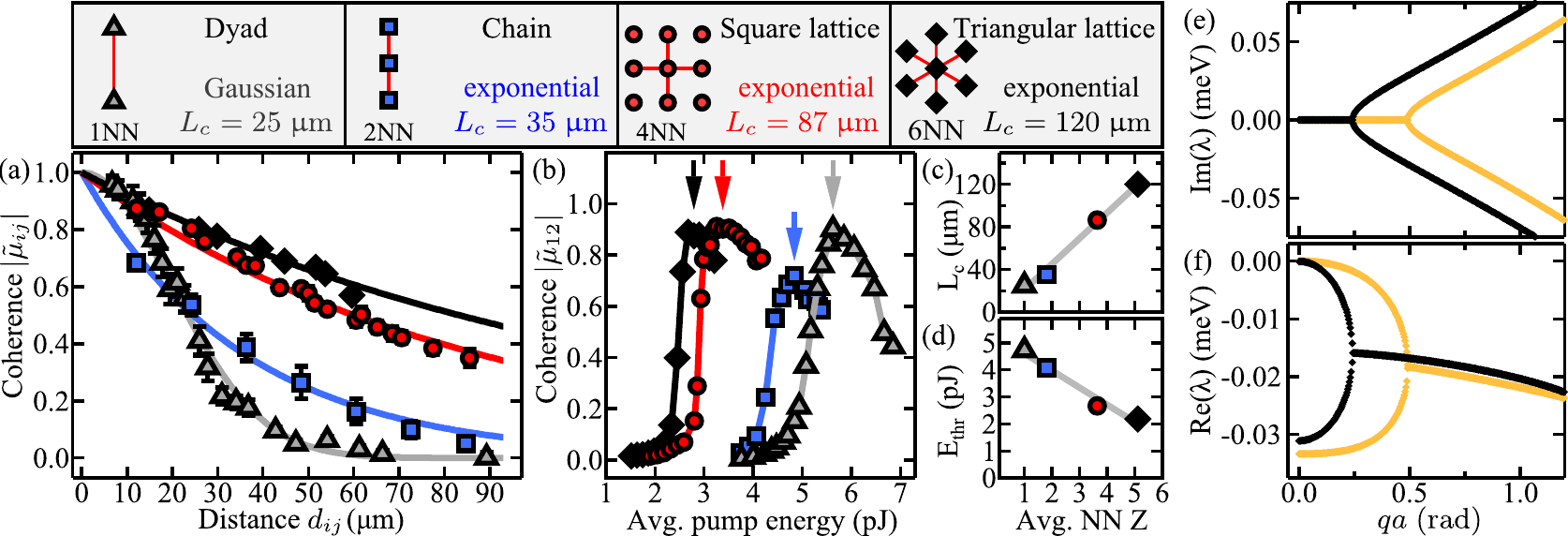}
	\caption{Comparison of coherence between ballistically coupled polariton condensates in different networks presented in Figs.~\ref{Fig1},~\ref{Fig3},~\ref{Fig4} and~\ref{Fig5}. (a) Spatial decay of coherence for a two condensate system of varying size (grey triangles), a 1D chain (blue squares), a square lattice (red squares) and a triangular lattice (black diamonds). Curves represent Gaussian and exponential fits to the experimental data points. (b) Excitation pulse energy dependence of the mutual coherence factor $| \tilde{\mu}_{12} |$ between the central-most condensate (index $1$) and one of its NNs (index $2$) for the 1D and 2D periodic structures, and between two condensates at $d_{12}=12.7\;\mathrm{\upmu m}$ for the case of a dyad. The pump powers used in (a) are marked with vertical arrows and are given by $P\approx 1.2P_\mathrm{thr}$, where $P_{\mathrm{thr}}$ is the threshold pump power for each system, respectively. (c,d) Coherence length $L_c$ and average excitation pulse energy $E_{\mathrm{thr}}$ per condensate node at threshold versus the average number of NNs $Z$. (e,f) Calculated Lyapunov exponents in $\Gamma \to X$ crystal momentum space ($q$-space) around the $\Gamma$-point ($qa=0$) belonging to a square condensate lattice with small (high) particle numbers given by yellow (black) curves.}
	\label{Fig6}
\end{figure*}

To compare the obtained results, we summarise in Fig.~\ref{Fig6}(a) the spatial decay of coherence for four different types of networks; two coupled condensates, 1D chain, and 2D square and triangular lattices. For the lattices the data points for different separation distances $d_{ij}$ correspond to different pairs of condensate nodes $\{i,j\}$. For the polariton dyad the physical separation distance between the two condensate nodes is changed. In all cases the pump power was chosen such as to maximise the coherence $|\tilde{\mu}_{12}|$ between a pair of NNs in the system. In Fig.~\ref{Fig6}(c) we illustrate the extracted (effective) coherence length $L_c$ versus the average number of NNs in each system.  It is apparent that the coherence length $L_c$ is increasing for systems with larger connectivity which we explain in the following. Interestingly, the mutual coherence between NNs in the 1D chain ($d_{ij} = a$) is noticeably lower than in the other systems. Perhaps most surprisingly, the chain's coherence is lower than the dyad for small $d_{ij}$ and then overtakes it around $d\gtrsim 25\;\mathrm{\upmu m}$.

We attribute the origin of increased spatial coherence for condensate networks with larger connectivity to their reduced population of reservoir excitons as we argue in the following. In Fig.~\ref{Fig6}(b) we compare the dependence of coherence $|\tilde{\mu}_{12}|$ between a pair of NNs as a function of the average excitation pulse energy per condensate in each system. We observe a reduction of threshold pulse energy $E_{\mathrm{thr}}$ per condensate node for networks with larger connectivity (larger number of NNs), which is shown in Fig.~\ref{Fig6}(d) and attributed to the increased gain given by ballistic exchange of particles between neighbouring nodes. A lower pump energy generates a smaller reservoir of excitons for each condensate node and since reservoir-condensate interactions play a dominant role in condensate dephasing~\cite{Wouters_PRB2009}, a lower threshold pump energy will generally result in increased coherence properties.

By discretising the system into a set of interacting condensates~\cite{topfer2020time} one can show that threshold pump power and coherence length scale approximately linear with the number of condensate NNs. Let us consider the steady state (without phase frustration) of frequency $\nu$ such that the discretised Gross-Pitaevskii equation can be written,
\begin{equation}
\nu \psi_n = \left[ V + \frac{i}{2} \left( \frac{ P}{n_s +  |\psi_n|^2} - \gamma\right) \right] \psi_n + (\epsilon + i \kappa ) \sum_{\langle nm \rangle} \psi_m,
\end{equation}
Here, $\psi_n$ describes the phase and amplitude of the $n$th with sum over NNs, the potential $V$ describes blueshift coming from the condensate interactions with itself and its noncondensed particle background, $P$ is the laser power, $n_s$ is the condensate saturation density, and $\gamma$ is the polariton lifetime. The non-Hermitian interaction between condensates is captured with $\epsilon$ and $\kappa$. The condensation threshold (the lasing mode) belongs to the mode with the lowest particle decay and its background reservoir of noncondensed particles responsible for dephasing is written,
\begin{equation} \label{eq.threshold_reservoir}
n_x^\text{(thr)} = n_s - \frac{Z \kappa}{R},
\end{equation}
where $Z$ is the number of NNs and $R$ is the scattering rate of reservoir particles into the condensate. The density of the reservoir $n_x^\text{(thr)}$ is proportional to the amplitude of random fluctuations experienced by the condensate (see Methods) and therefore its coherence properties will depend on the number of neighbours $Z$. The first order spatial correlation function can be written,
\begin{equation}
\mu_{nm}  \propto \exp{[-f_{\mathbf{k},\omega}(\mathbf{r}_{nm})]},
\end{equation}
where $\mathbf{r}_{nm}$ is the distance vector between condensates $n$ and $m$ and $f_{\mathbf{k},\omega}$ is the phase-phase correlation function which determines the condensate's long range coherence properties~\cite{Szymanska_PRL2006}. It is beyond the scope of this study to calculate $f_{\mathbf{k},\omega}$ given the complicated spatial density of the condensate. At the length scale of the experiment we observe an exponential decay of spatial correlations in all cases (except for the polariton dyad) shown in Fig.~\ref{Fig6}(a) and therefore we will assume that $f_{\mathbf{k},\omega} = |\mathbf{r}_{nm}|/L_c$ holds. Assuming frequency independent noise processes the coherence length $L_c$ becomes inversely proportional to the strength of the noise~\cite{Szymanska_PRL2006, Roumpos_PNAS2012} and one obtains,
\begin{equation} \label{eq.coherence_length}
L_c \propto \frac{1}{\gamma +  Rn_s - Z \kappa} \simeq \frac{1}{\gamma+Rn_s} \left[ 1 + \frac{Z \kappa}{\gamma + R n_s} \right].
\end{equation}
The expected linear dependencies of both threshold pump power $P_\mathrm{thr} \propto n_x^\text{(thr)}$ (Eq.~\eqref{eq.threshold_reservoir}) and coherence length $L_c$ (Eq.~\eqref{eq.coherence_length}) are both indicated as gray lines in Figs.~\ref{Fig6}(c,d).
We point out that similar results (increasing coherence with larger connectivity and dimensionality in networks of coupled elements) have been observed in other technological platform such as in arrays of coupled VCSELs~\cite{Golshani_JoAP_1999,pier2000strain}, micromechanical oscillator arrays~\cite{Zhang_PRL2015} as well as for coupled fiber lasers~\cite{Fridman:s}.\\

It is instructive to investigate the condensate's dispersion of elementary excitations (Lyapunov exponents) through the GPE which directly relates to the behaviour of space-time correlations in the system. For simplicity we will focus on the case of a 2D square lattice such as displayed in Fig.~\ref{Fig5} in a steady state (continuous wave excitation). For our calculation of the Lyapunov exponents we will assume that we are working in the the bulk of the condensate and therefore the system is taken to have discrete translational invariance. This allows us to apply Bloch's theorem to the standard Bogoliubov treatment and solve the Lyapunov exponents in the reduced Brillouin zone of the condensate-pump lattice (see Appendix~\ref{appendix_bog}). In Fig.~\ref{Fig6}(e,f) we plot the Lyapunov exponents around the $\Gamma$-point of the reduced Brillouin zone for two different numbers of particles in the condensate. Yellow and black coloured curves correspond to $N\approx 100$ and $\approx 800$ particles in the condensate unit cell respectively for a lattice constant of $a = 12$ $\upmu$m. $\text{Im}{(\lambda)}$ corresponds to oscillatory evolution of the fluctuations, showing the purely diffusive Goldstone branch [$\text{Im}{(\lambda)}=0$] at lower momenta. $\text{Re}{(\lambda)}$ corresponds to the decay (growth) rate of fluctuations when negative (positive) valued. In the top branch of the pitchfork, we observe that $\text{Re}{(\lambda)}$ becomes more negative for a given wavevector as more particles are in the condensate (black curve). This is in agreement with our experiment where coherence increases with pump power (particle number grows and decay of fluctuations increases) up to the point where single-mode (stationary) behaviour is lost. Our results are similar to those obtained for spatially uniform systems~\cite{Szymanska_PRL2006, Wouters_PRL2007, Chiocchetta_EPL2013} underlining that spatial details of the condensate structure are not relevant to long-wavelength fluctuations.

\section{Discussion}
We have demonstrated that the coherence properties in lattices of polariton condensates are enhanced by balancing the condensate emission across the system using a closed-loop feedback scheme to adjust the excitation pump geometry. This scheme reduces the effects of optical aberrations in the experimental system, as well as counteracts mode localisation due to sample non-uniformities~\cite{Pier:97} and non-homogeneous gain distribution across the coupled condensate network. While actively controlling the condensate lattice uniformity we have accurately determined phase and coherence between any pairs of condensates in different types of networks ranging from two condensate-systems to 1D and 2D periodic structures. The dynamics and synchronisation of coupled non-linear elements is critically influenced by the underlying coupling topology~\cite{watts_collective_1998}. Here, we have shown that an increase in connectivity, i.e. number of NNs, significantly reduces the operational pump power per node and increases the coherence length demonstrating a promising route to polaritonic devices with networks of many coupled condensates with potential application in simulators and optical based computation. The presented measurements and techniques provide a deeper understanding on the coherence properties of coupled light-matter wave fluids in  low dimensional quantum systems, and qualify for other open (dissipative) networks such as laser arrays and photon condensate lattices. Furthermore, our analysis of the lattice condensate fluctuations reveals similar long-wavelength dispersions to those of uniform systems which are regarded as the ideal case from a theoretical point of view. This suggests access to fundamental long-wavelength physics belonging to uniform systems by designing instead an extended condensate lattice.

\section{Acknowledgements}
The authors acknowledge the support of the Skoltech NGP Program (Skoltech-MIT joint project), the UK’s Engineering and Physical Sciences Research Council (grant EP/M025330/1 on Hybrid Polaritonics), and the RFBR projects No. 20-52-12026 (jointly with DFG) and No. 20-02-00919. Y.G.R. acknowledges support by CONACYT (Mexico) Grant No.\ 251808 and by PAPIIT-UNAM Grant No.\ IN106320.
\section{Methods}
\subsection{Experimental methods}
We utilise a strain-compensated $2\lambda$ GaAs microcavity with embedded InGaAs quantum wells and a quality factor $Q\approx 12\;000$~\cite{cilibrizzi2014polariton} . We continuously cool the microcavity sample in a cold-finger cryostat ($T\approx 6\;\mathrm{K}$) and operate at a negative cavity-detuning of $\Delta \approx -5 \;\mathrm{meV}$ resulting in a lower polariton mode at $\lambda\approx 858\;\mathrm{nm}$ for zero in-plane momentum $|\mathbf{k}| =0$. We use a circularly polarised non-resonant blue-detuned ($\lambda \approx 800\; \mathrm{nm}$) pulsed laser (pulse duration $\approx 150\;\mathrm{fs}$, repetition rate $80\;\mathrm{MHz}$). For each condensate node the laser is focused onto the microcavity sample to a beam waist of approximately $2\;\mathrm{\upmu m}$ via a $0.4\;\mathrm{NA}$ microscope objective. Polariton photoluminescence (PL) and the excitation laser are measured by imaging the cavity emission or reflected light of the microcavity sample onto a CCD sensor with integration times in the range of $10$-$100\;\mathrm{ms}$, which corresponds to averaging measurements over approximately one million realisations of the system. By using a longpass or shortpass filter in front of the camera we can selectively choose to measure either the spatial geometry of the polariton emission or the excitation laser.

Experimental details of the density stabilisation technique for polariton condensate lattices are given in Appendix~\ref{appendix_FeedbackScheme}. Methods for the measurement of the mutual complex coherence factor $\mu_{ij}$ between any pair of condensate nodes are described in Appendices~\ref{appendix.FarfieldMeasurementTechnique} and~\ref{appendix.MeasurementOfCoherence}.

\subsection{Error analysis}
Error bars for the absolute values of the complex coherence factor $|\mu_{ij}|$ represent the confidence interval of the fitted parameter and are extracted as described in Ref.~\cite{Singer:12}. We fit the double hole interference pattern (Eq.~\ref{eq.FarfieldInterference2Spots_1D}) to the experimental interference pattern $I_{1+2}$ with three fit parameters $|\mu_{ij}|$, $|\vect{d}_{12}|$ and $\theta_{12}$. For each measurement we define the quality of the fitted interference curve $I_\mathrm{fit}$ by the dimensionless quantity
\begin{equation*} \label{Eq.FitQuality}
R =  \frac{  \sum_{k_{||}} \left[ I_\mathrm{fit}(k_{||}) -   I_{1+2}(k_{||})  \right]^2 }{ \sum_{k_{||}}  \left[  I_{1+2}(k_{||}) \right] ^2}
\end{equation*}
where summation is performed over all sample points $k_{||}$. Next, we keep all fit parameters constant except the coherence factor $|\mu_{ij}|$ and recalculate the quality factor $R$ for varying $|\mu_{ij}|$. The confidence interval of $|\mu_{ij}|$ is determined as the range in which the quality factor $R$ increases by $100\%$ from its minimum value.

\subsection{Numerical Simulations}
A stochastic dissipative Gross-Pitaevskii equation describes the polariton condensate order parameter $\Psi(\mathbf{r},t)$ coupled to a rate equation describing the density of a background excitonic reservoir $n_x$ feeding particles into the condensate. The reservoir itself is sustained by a decaying population of excited electron-hole pairs $n_c(\mathbf{r},t) = P(\mathbf{r}) e^{-(W + \Gamma_c)t}$ which we assume are generated instantaneously by the sub-picosecond nonresonant Gaussian shaped pump $P(\mathbf{r})$,
\begin{subequations} \label{eq.GP_Res}
\begin{align} \label{eq.GP}
i  \frac{\partial \Psi}{\partial t} & = \frac{1}{2}\left[  - \frac{\hbar\nabla^2}{m} + V+ \alpha |\Psi|^2 + i \left( R n_x - \gamma \right) \right] \Psi + \frac{dW}{dt},\\ \label{eq.ResA}
\frac{\partial n_x}{\partial t} & = - \left( \Gamma_R + R |\Psi|^2 \right) n_x + W n_c.
\end{align}
\end{subequations}
Here, $m$ is the effective mass of a polariton in the lower dispersion branch, $\alpha$ is the interaction strength of two polaritons in the condensate, $g$ is the polariton-reservoir interaction strength, $R$ is the rate of stimulated scattering of polaritons into the condensate from the active reservoir, $\gamma$ is the polariton condensate decay rate, $\Gamma_{R,c}$ are the decay rates of the reservoir excitons and electron-hole pairs respectively, $W$ is the conversion rate between the reservoir excitons and electron hole-pairs, and
$V(\mathbf{r},t) = g(n_x(\mathbf{r},t)+n_c(\mathbf{r},t))$
is the pump-induced potential with an effective reservoir-condensate interaction strength $g$. Here we have introduced a Gaussian white noise term $dW$ based on a Monte Carlo technique in the truncated Wigner representation~\cite{Wouters_PRB2009}. The correlators are written $\langle dW(\mathbf{r}_i) dW(\mathbf{r}_j) \rangle  =0 $ and $\langle dW^*(\mathbf{r}_i) dW(\mathbf{r}_j) \rangle  = (\Gamma + R n_x) \delta(\mathbf{r}_i - \mathbf{r}_j) dt/2\Delta A$ where $\Delta A$ is the cell area of the spatial grid. The coherence factor (Eq.~\ref{eq.integratedcomplexcoherencefactor}) is then calculated by averaging over multiple realisations of condensate formation. Good agreement between simulation and experimental data is obtained by only adjusting the values of $g$ and $\text{max}[P(\mathbf{r})]$ between different systems. The reason $g$ is taken as an tuneable parameter is due to the fact that $V$ can possess a more complicated sublinear dependence in Eq.~\eqref{eq.GP} as pointed out previously~\cite{Gao_APL2015}. Parameters were set to: $\Gamma = 1/5.5$ ps$^{-1}$, $\Gamma_R = 2\Gamma$, $\Gamma_c = 0.0055 \Gamma$, $W = 0.275 \Gamma$, $\alpha = 0.01$ ps$^{-1}$ $\upmu$m$^2$, $R  =0.05 \alpha$, $m = 0.32$ meV ps$^2$ $\upmu$m$^{-2}$, and $g = \{ 0.0008,\ 0.0011,\ 0.0022, \ 0.0016\} \alpha$ for dyad, chain, square, and triangle respectively.

\appendix
\renewcommand{\appendixname}{APPENDIX} 
\renewcommand{\thesection}{\Alph{section}} 
\section{COHERENCE IN POLARITON LATTICES} \label{appendix_Coherence}
In our study of synchronisation in networks of coupled polariton condensates we are interested in the correlations between pairs of condensates which are described by their mutual coherence function
\begin{equation*}
\Gamma_{ij}(t) = \left\langle \psi_i(t)^*  \psi_j(t) \right\rangle,
\end{equation*}
where the brackets denote ensemble averaging and $\psi_i(t)$ is the complex-valued amplitude of the $i$th condensate node. The expectation value of the condensate's particle number (occupation) is given by the diagonal elements $\Gamma_{ii}(t)$, and the presence of non-zero off-diagonal elements $|\Gamma_{ij}(t)| > 0$ for $i \neq j$ indicates long-range order in the coupled condensate network. For nonstationary systems (such as for polariton condensates generated under pulsed excitation) the coherence function $\Gamma_{ij}(t)$ depends on time $t$. We note that pulsed excitation ($\approx 150\;\mathrm{fs}$ pulse width) of ballistically coupled condensate nodes leads to polariton photoluminescence with typical emission signal full-width-at-half-maximum (FWHM) of $18\;\mathrm{ps}$ (see Appendix~\ref{appendix_TimeResolvedEmission}). A normalised form of the mutual coherence function is given by the complex coherence factor
\begin{equation*}
\mu_{ij}(t) =\frac{   \Gamma_{ij}(t) }{ \sqrt{  \Gamma_{ii}(t)   \Gamma_{jj}(t)   }},
\end{equation*}
where for a fully coherent pair of condensates $|\mu_{ij}|$ = 1. Considering measurements with time integration over many pulses we define the integrated complex coherence factor
\begin{equation*}
\tilde{\mu}_{ij} =\frac{  \int \Gamma_{ij}(t) \mathrm{d}t }{ \sqrt{  \int  \Gamma_{ii}(t)  \mathrm{d}t   \int  \Gamma_{jj}(t)  \mathrm{d}t } },
\end{equation*}
as an averaged measure for the mutual coherence properties of condensates in polariton networks. Throughout this manuscript time-integrated variables are marked with the tilde diacritic. For stationary ergodic systems (such as polariton condensates under constant pumping) the complex coherence factor $\mu_{ij}(t)$ does not depend on time and time-integrated measurements of correlations fully determine the coherence properties, i.e. $\mu_{ij}(t) = \tilde{\mu}_{ij}$. In case of pulsed excitation of polariton condensates the modulus of the time-integrated factor $|\tilde{\mu}_{ij}|$ represents a lower bound for the maximum value of the complex coherence factor $|\mu_{ij}(t)|$. We note that the argument of the integrated complex coherence factor represents the average condensate node phase difference $\tilde{\theta}_{ij} = \arg(\tilde{\mu}_{ij})$ and the integrated particle number $\int \Gamma_{ii}(t)  \mathrm{d}t$ of each condensate node is proportional to its measurable average emission intensity $\tilde{I}_i$.
\section{FEEDBACK SCHEME} \label{appendix_FeedbackScheme}
In this manuscript we present networks of ballistically coupled polariton condensates for which the pumping geometry has been adjusted using a reflective SLM (see Fig.~\ref{Fig1}) such as to equalise the emission power of each condensate node at condensation threshold. The algorithm used to calculate the holograms which are applied to the SLM is a modified version of the Gerchberg-Saxton (GS) algorithm~\cite{gerchberg1972practical} and takes into account feedback from the recorded PL. This iterative algorithm is similar to the schemes presented in Refs.~[\onlinecite{PhysRevX.4.021034},\onlinecite{tamura2016highly}] applied to trapping of cold atoms. In our case the recorded PL of the condensates is not directly proportional to the pump power of each node but depends on the coupling-topology of the network, i.e. ballistic coupling between spatially separated condensate nodes affects the gain of each condensate.

\begin{figure}[!t]
	\center
	\includegraphics[]{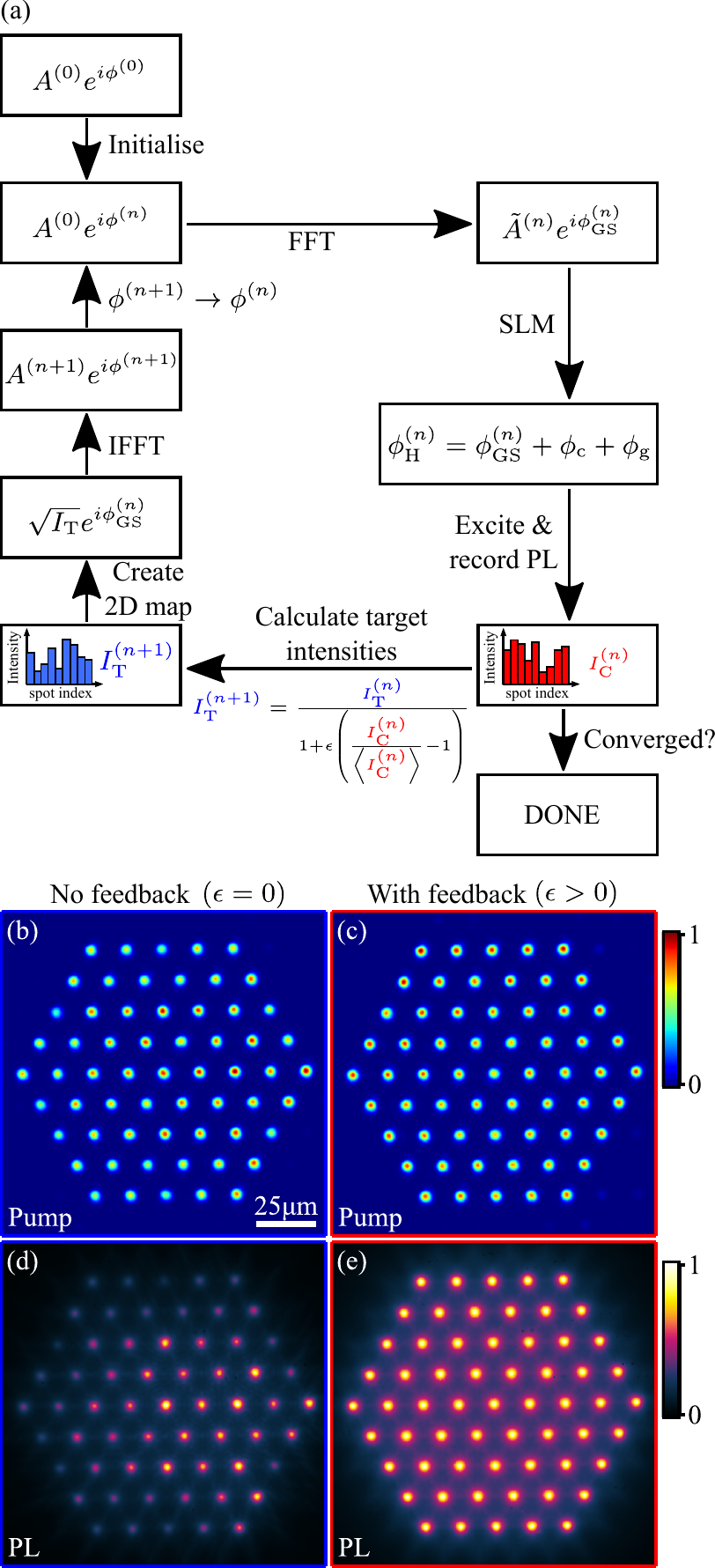}
	\caption{(a) Schematic of the modified iterative Gerchberg-Saxton algorithm including feedback of the polariton photoluminescence (PL) to stabilise the node density in networks of coupled polariton condensates. (b,c) Recorded laser pump profile and (d,e) real space PL at condensation threshold without and with feedback, respectively.}
	\label{Fig_SI_Algorithm}
\end{figure}

The algorithm is schematically illustrated in Fig.~\ref{Fig_SI_Algorithm}(a) and begins with the initialisation of a 2D complex field of amplitude $A^{(0)}(\mathbf{r})$ and phase $\phi^{(0)}(x,y)$ representing the amplitude of the pump laser and the phase hologram in pixel coordinates of the SLM, respectively. As an initial guess we assume a constant phase pattern $\phi^{(0)}=0$ and a Gaussian field amplitude, whose width matches the pump laser width. In each iteration step $n$ propagation of the complex field $A^{(0)} e^{i \phi^{(n)}(\mathbf{r})}$ from the SLM plane to the focal plane is computed by a fast Fourier transform (FFT) yielding a complex-valued target field with amplitude $\tilde{A}^{(n)}(\mathbf{r})$ and phase $\phi_{\mathrm{GS}}^{(n)}(\mathbf{r})$. We add the output phase pattern $\phi_{\mathrm{GS}}^{(n)}$ with a device dependent correction term $\phi_c$, which can also take into account optical aberrations in the experimental setup, and a blazed grating $\phi_g$ to offset the modulated laser beam from its unmodulated reflected part. The sum of these terms modulo the $2\pi$ bitlevel is applied to the SLM and defines the pump laser geometry. In a next step we record the near-field photoluminescence of the polariton system excited by the current pump laser geometry and extract the emission power $I_\mathrm{C}^{(n)}$ of each condensate by integrating the recorded signal within the FWHM ($\approx 2\;\mathrm{\upmu m}$) of each node. We update the target power $I_\mathrm{T}^{\mathrm{(n+1)}}$ for the next iteration step of the algorithm for each condensate node utilising the non-linear function
\begin{equation}
\label{eq.FeedbackNonlinearFunction}
I^{(n+1)}_T = \frac{I^{(n)}_T}{1+\epsilon\left(I^{(n)}_C /  \langle I^{(n)}_C \rangle  -1  \right)},
\end{equation}
where $\epsilon$ is an adjustable feedback parameter and $\langle I^{(n)}_C \rangle$ denotes the mean of the measured condensate node emission distribution. The non-linear mapping given by Eq.~\eqref{eq.FeedbackNonlinearFunction} lowers (raises) the target pump power $I^{(n+1)}_T$ of nodes with measured emission power $I^{(n)}_C$ larger (lower) than the mean value $\langle I^{(n)}_C \rangle$. The parameter $\epsilon$ controls the speed of the feedback loop but cannot be set too large to avoid destabilisation of the algorithm. For $\epsilon = 0$ the algorithm represents the conventional form of the GS algorithm with no iterative adjustment of target spot intensities, i.e. $I^{(n+1)}_T = I^{(n)}_T$. Next, a complex-valued target field with amplitude $\sqrt{I_\mathrm{T}}(\mathbf{r})$ and phase $\phi_{\mathrm{GS}}^{(n)}(\mathbf{r})$ is constructed. Hereby, the field amplitude $\sqrt{I_\mathrm{T}}(\mathbf{r})$ consists of the superposition of Gaussian spots representing the position of each condensate node. The amplitude of each Gaussian peak is given by the square root of the updated spot power, i.e. $\sqrt{I_\mathrm{T}^{\mathrm{(n+1)}}}$ and the width is related to the width of the Gaussian profile of the pump laser by means of a Fourier transform. Back-propagation of this complex field to the SLM plane is computed by an inverse fast Fourier transform (IFFT) yielding a phase pattern $\phi^{(n+1)}$ which replaces the initial phase pattern for the next iteration step, i.e. $\phi^{(n+1)} \rightarrow \phi^{(n)}$. The amplitude field at the SLM plane is kept constant as the initial field $A^{(0)}$ for each iteration step. We quit execution of the algorithm once the measured spot power distribution $I_\mathrm{C}^{(n)}$ shows a spread which is smaller than a certain threshold level (usually $1\%\;\mathrm{(RSD)}$ reached within $<100\;$iterations).\\

We find good results for stabilising the intensity distribution in ballistically coupled polariton networks ($\mathrm{RMS} \leq 1\%$) pumped at condensation threshold $P \gtrsim P_{\mathrm{thr}}$ by choosing $\epsilon$ to be in the range $10^{-2} - 10^{-1}$. Stabilisation at larger pump powers $P \gg P_{\mathrm{thr}}$ is impeded using the presented technique - which operates at a typical rate of $60\;\mathrm{Hz}$ - because of stronger nonlinear effects resulting in unstable regimes~\cite{Wouters_PRL2007, Bobrovska_PRB2014, Baboux_Optica2018} and/or excitement of polaritons to higher energy states~\cite{Wertz_NatPhys2010}, with subsequent nonstationary condensate dynamics on a picosecond timescale. \\

In Fig.~\ref{Fig_SI_Algorithm}(b) we show the recorded laser (pump) profile for a triangular lattice of 61 nodes (lattice constant $a=14.9\;\mathrm{\upmu m}$) using a phase hologram computed with the conventional GS algorithm ($\epsilon=0; 100\;\mathrm{iterations}$). Although the target image is a laser pattern with homogeneous spot intensities, due to limited accuracy of the algorithm as well as unavoidable optical aberrations due to misalignment and device imperfections, the resultant experimentally obtained pump spot intensity distribution deviates from the homogeneous target. We measure a spread of $\approx 17\%$ (RSD) for the distribution of pump spot powers. Using the same pump profile to excite a triangular lattice of polariton condensates at condensation threshold ($P\approx P_{\mathrm{thr}}$) we record the near-field PL as shown in Fig.~\ref{Fig_SI_Algorithm}(d) and find an increased spread of condensate node densities of $\approx 37\%\;\mathrm{(RSD)}$. While sample disorder can have an impact on the spatially dependent condensation threshold pump power and emission intensity for each condensate node, the main reason for the increased spread in the condensate emission power distribution is due to the system's nonlinear input-output-power characteristics and the finite size of the system. Nodes that are positioned at the edge of the condensate network are coupled to fewer nearest-neighbours (NN) than nodes in the bulk. Since coherent coupling between ballistically expanding polariton condensates results in a reduced condensation threshold~\cite{Cristofolini_PRL2013}, nodes coupled to fewer NNs will typically have a higher threshold pump power than nodes with a greater number of NNs  (see section~\ref{section_dimensionality}). As a result, condensate nodes in the bulk of the network generally have a larger occupation number (emission) than condensate nodes at the edges when pumped with the same excitation pump power~\cite{kalinin2018networks}. This is a consequence of polariton waves radiating to the outside continuum and escaping the network which can be regarded as an effective flux-induced potential~\cite{Ostrovskaya_PRL2013}.\\
In Figs.~\ref{Fig_SI_Algorithm}(c,e) we show the recorded pump laser and corresponding near-field polariton emission profiles at threshold $P\approx P_{\mathrm{thr}}$ after applying the intensity stabilisation feedback loop and terminating it with spread in condensate node densities of $\approx 1\%\; \mathrm{(RSD)}$.  Comparison of the density spread and the spatial coherence in both polariton lattices without ($\epsilon = 0$) and with ($\epsilon > 0$) is shown in Fig.~\ref{Fig1}.
\section{FARFIELD INTERFERENCE OF POLARITON LATTICES} \label{appendix.FarfieldMeasurementTechnique}
The time-averaged far-field interference pattern $\tilde{I}(\mathbf{k})$ of a set of $N$ partially coherent, narrow-bandwidth and point-like light sources positioned in one plane can be written in the basis of spatial frequencies $\mathbf{k}$ as (see Appendix~\ref{appendix_farfield_coherence})
\begin{equation}
\label{eq.SchellsDiscretisedTheorem}
\tilde{I}(\mathbf{k}) \propto \sum_{i,j =1}^{N} \sqrt{\tilde{I_i}\tilde{I_j} } \tilde{\mu}_{ij} \mathrm{e}^{i \mathbf{k} \cdot \mathbf{d}_{ij}}.
\end{equation}
For each pair $\{i,j\}$ of point sources $\mathbf{d}_{ij}$ denotes their in-plane spatial separation vector and $\tilde{\mu}_{ij}$ their respective integrated complex coherence factor. For an incoherent system ($\tilde{\mu}_{ij} = 0$ for $i \neq j$) the resulting homogeneous far-field radiation pattern is the incoherent superposition of light sources with individual intensities $\tilde{I}_i$. However, for non-vanishing off-diagonal coherence elements ($|\tilde{\mu}_{ij}|>0$) the radiation pattern $\tilde{I}(\mathbf{k})$ described in Eq.~\eqref{eq.SchellsDiscretisedTheorem} becomes inhomogeneous and is formed by the sum over discrete Fourier-components $\exp{(i \mathbf{k} \cdot \mathbf{d}_{ij})}$ weighted by the integrated complex coherence factor $\tilde{\mu}_{ij}$.\\

\begin{figure}[!t]
	\center
	\includegraphics[]{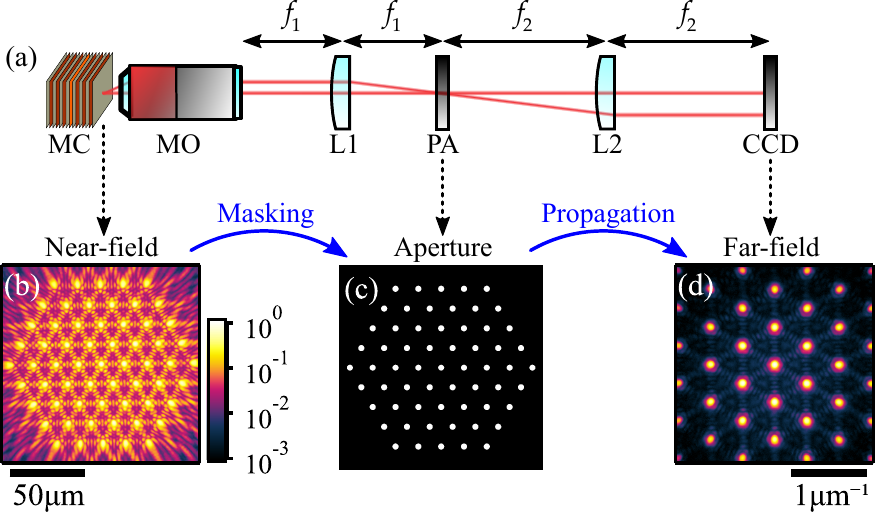}
	\caption{Farfield interference of coherently coupled condensates. (a) Schematic of the experimental detection setup with microcavity (MC), microscope objective lens (MO), lenses (L1,L2), programmable aperture (PA) and charge-coupled device (CCD). (b) Real space PL, (c) real space masking aperture and (d) corresponding far-field interference pattern of 61 condensates arranged in a triangular geometry.}
	\label{Fig_FarfieldMeasurement}
\end{figure}

Hence, in analogy to beam interference measurements of coupled laser arrays~\cite{Mahler_PRL2020}, we can investigate the far-field emission of the coupled polariton condensate network to gain information about the system's spatial coherence properties. To remove residual PL coming from polaritons outside the condensation (gain-)centres we spatially filter the emission of each condensation center and measure the interference of the masked condensate emission in the far field. As schematically illustrated in Fig.~\ref{Fig_FarfieldMeasurement}(a) we project the near-field PL onto a programmable aperture which allows to selectively mask the real space field $\Psi(\mathbf{r},t) \rightarrow A(\mathbf{r}) \Psi(\mathbf{r},t)$ with amplitude function $A(\mathbf{r}) \in (0,1)$. Subsequently we image the diffraction pattern (far field) onto a charge-coupled device (CCD) sensor. The programmable aperture consists of a digitally controllable reflective liquid-crystal SLM as described in reference~\cite{Lousberg:06}. With the use of an additional wave plate and linear polariser the apparatus transmits only circular polarised light which is chosen to be the same as the polarisation of the pump laser beam. By projecting the near-field PL of the polariton condensate lattice (Fig.~\ref{Fig_FarfieldMeasurement}(b)) onto an aperture that transmits only the central $\mathrm{FWHM}\approx 2\;\mathrm{\upmu m}$ of each condensation centre (Fig.~\ref{Fig_FarfieldMeasurement}(c)) we record the corresponding far-field emission (Fig.~\ref{Fig_FarfieldMeasurement}(d)) that consists of the interference of all condensate nodes. According to the Fraunhofer diffraction formula the measured far-field intensity distribution for a fully coherent wave is proportional to the squared modulus of the Fourier transform of the optical field at the aperture location, which is obtained from Eq.~\eqref{eq.SchellsDiscretisedTheorem} for $|\tilde{\mu}_{ij}| =1$ for all pairs $\{i,j\}$. Indeed, the appearance of the reciprocal triangular lattice, shown in Fig.~\ref{Fig_FarfieldMeasurement}(d), indicates long-range coherence across the triangular lattice of polariton condensates.
%
\section{MEASUREMENT OF COHERENCE IN POLARITON LATTICES} \label{appendix.MeasurementOfCoherence}
\begin{figure*}[!t]
	\center
	\includegraphics[]{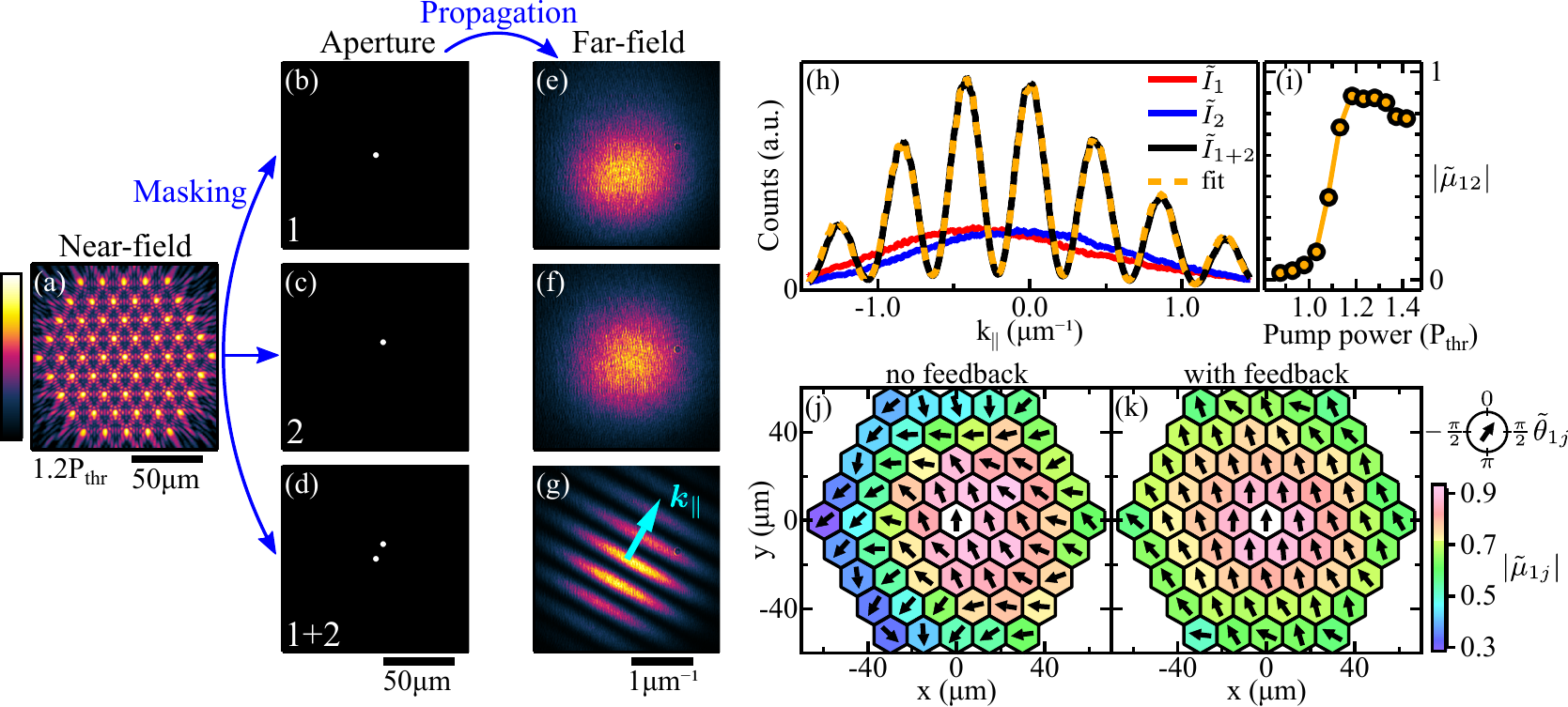}
	\caption{Measurement of spatial coherence in lattices of coupled polariton condensates. By projecting the (a) real space PL of a lattice of coupled condensates onto a programmable aperture that spatially filters the emission to transmit only (b) condensate node $1$, (c) condensate node $2$ or (d) both condensate nodes $1+2$ we record the respective far-field emission (e-g). The extracted intensity profiles $\tilde{I}_1(k_{\parallel})$, $\tilde{I}_2(k_{\parallel})$ and $\tilde{I}_{1,2}(k_{\parallel})$ perpendicular to the orientation of the interference fringes in (g) are shown in (h) together with a fit according to Eq.~\eqref{eq.FarfieldInterference2Spots_1D}. Pump power-dependence of the modulus of the integrated coherence factor $|\tilde{\mu}_{12}|$ is displayed in (i). Extracted values of the integrated complex coherence factor $\tilde{\mu}_{1i}$ between the central condensate node $1$ and each other condensate node $i$ are illustrated in (j) and (k) for the cases of no condensate node density stabilisation and with node density stabilisation. All values are extracted at a total pump power $P=1.2P_{\mathrm{thr}}$. False colour scale shown for (a) applies to (e-g) in linear scale and to (a) in logarithmic scale saturated below $0.01$ of the maximum count rate. }
	\label{Fig_CoherenceMeasurement}
\end{figure*}

Far field measurements shown in Fig.~\ref{Fig_FarfieldMeasurement} give information about the global coherence properties of the coupled condensate network and are analogous to time-of-flight experiments used in cold atom systems~\cite{Hadzibabic_Nature2006}. However, precise information about the local complex coherence factor $\mu_{ij}$ can not be obtained. The programmable aperture in our experiment (see Fig.~\ref{Fig_FarfieldMeasurement}) overcomes this problem and allows for selective interference between any pair $\left\{ i,j \right\}$ of condensates, such that the complex coherence factor $\mu_{ij}$ can be spatially reconstructed~\cite{Lousberg:06, Lundeberg_APL2007}. In particular, the far-field interference of the emission of two spatially filtered condensate nodes $\psi_i(\vect{r},t)$ and $\psi_j(\vect{r},t)$ centred around their respective positions $\vect{r}_i$ and $\vect{r}_j$ yields the measurable intensity pattern $\tilde{I}_{i+j}(\vect{k})$ in $\vect{k}$-space as,~\cite{Thompson:57}
\begin{equation}
\begin{aligned}
\label{eq.FarfieldInterference2Spots}
\tilde{I}_{i+j}(\vect{k}) &= \tilde{I}_i(\vect{k}) + \tilde{I}_j(\vect{k}) \\
&\quad + 2\sqrt{\tilde{I}_i(\vect{k}) \tilde{I}_j(\vect{k}) } \left| \tilde{\mu}_{ij} \right| \cos\left(\vect{k} \cdot \vect{d}_{ij} + \tilde{\theta}_{ij} \right),
\end{aligned}
\end{equation}
where we explicitly take into account non-homogeneous intensity distributions $\tilde{I}_{i,j}(\vect{k})$ due to the finite aperture sizes. In Fig.~\ref{Fig_CoherenceMeasurement} we project the triangular lattice of 61 condensates (a) onto three different apertures filtering the emission of the central-most condensate (b), of one of its nearest-neighbours (c) and of both condensates simultaneously (d). The corresponding measured far-field diffraction patterns $\tilde{I}_1(\vect{k})$, $\tilde{I}_2(\vect{k})$ and $\tilde{I}_{1+2}(\vect{k})$ are shown in Figs.~\ref{Fig_CoherenceMeasurement}(e-g) respectively. We extract 1D intensity profiles $\tilde{I}( k_{\parallel})$ along the direction vector $\vect{k}_{\parallel}$, which is defined as co-parallel to $\vect{d}_{12}$ such that $\vect{k}_\parallel \cdot \vect{d}_{12} = k_{\parallel} |\vect{d}_{12}|$. Equation~\eqref{eq.FarfieldInterference2Spots} can be rewritten as,
\begin{equation}
\begin{aligned}
\label{eq.FarfieldInterference2Spots_1D}
\tilde{I}_{1+2}(k_{\parallel}) &=\tilde{I}_1(k_{\parallel}) + \tilde{I}_2(k_{\parallel}) \\
			 &\quad  + 2\sqrt{\tilde{I}_1(k_{\parallel}) \tilde{I}_2(k_{\parallel}) } \left| \tilde{\mu}_{12} \right| \cos\left(k_{\parallel} |\vect{d}_{12}| + \tilde{\theta}_{12} \right),
\end{aligned}
\end{equation}
and is fitted to the experimentally extracted intensity profiles $\tilde{I}_1(k_{\parallel})$, $\tilde{I}_2(k_{\parallel})$ and $\tilde{I}_{1,2}(k_{\parallel})$ to yield $|\tilde{\mu}_{12}|$ and $\tilde{\theta}_{12}$ as shown in Fig.~\ref{Fig_CoherenceMeasurement}(h). By extracting $|\tilde{\mu}_{12}|$ for varying pump power we identify the threshold-like behaviour of coherence shown in Fig.~\ref{Fig_CoherenceMeasurement}(i) with vanishing coherence between the two nodes below threshold and a sharp increase at threshold. The observed decrease of coherence for larger pump powers $P>1.2P_\mathrm{thr}$ is attributed to effects involving dephasing due to increased particle interactions, reservoir induced noise, enhanced proliferation of topological defects, as well as to the emergence of multi-mode emission~\cite{Ohadi_PRX2016} reducing the fringe visibility in time-integrated measurements.

Repeating the interference measurement for the central condensate node $1$ and each other condensate node $j$ we fully characterise the spatial coherence properties of the central condensate node within the triangular lattice (see Figs.~\ref{Fig_CoherenceMeasurement}(j,k) or Figs.~\ref{Fig1}(e,f)).
\section{FINITE VS INFINITE LATTICE} \label{subsection_InfiniteLatticeSize}
\begin{figure}[!t]
	\center
	\includegraphics[]{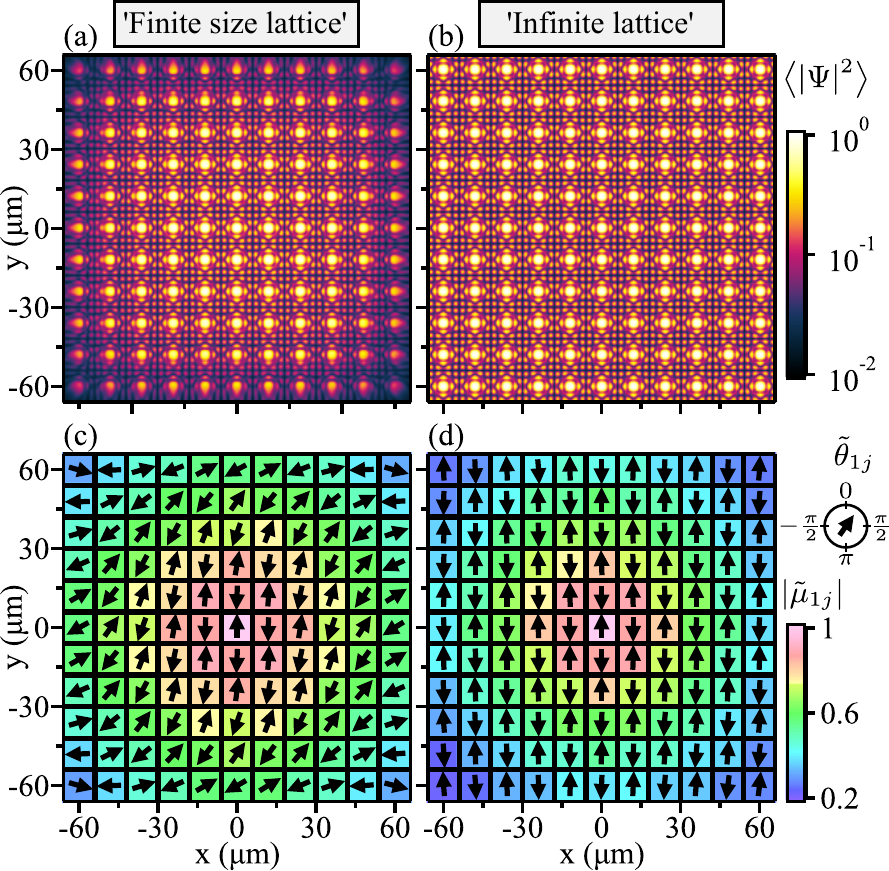}
	\caption{Numerical simulation of a square lattice of ballistically expanding polariton condensates with damped (a,c) and periodic boundary conditions (b,d). (a,b) Time-integrated and normalised real space densities for both types of lattices with equal pump power for each condensate node. (c,d) Extracted absolute values and phases of the complex degree of coherence $\mu_{1j}$ between the most central condensate node (index $1$, located at $x,y=0$) and every other condensate node (index $j$). The lattice constant is $a = 12$ $\upmu$m and size of simulation grid is $16a \times 16a$ with displayed results are zoomed in on an area of $11a \times 11a$.}
	\label{Fig_FiniteVsInfinite}
\end{figure}

In experiment finite-size effects cannot be avoided whereas in theory they can by applying periodic boundary conditions. In Fig.~\ref{Fig_FiniteVsInfinite} we investigate the difference of finite and periodic systems through simulation of Eq.~\eqref{eq.GP_Res}. In agreement with experiment, and previous theoretical observations~\cite{kalinin2018networks}, the pseudo-spin of $\tilde{\mu}_{ij}$ tilts as one approaches the edges of the finite (damped boundary condition) condensate lattice (see Fig.~\ref{Fig_FiniteVsInfinite}(c)). This effect is expected due to the non-zero flux of particles through the boundaries of the lattice resulting in a phase gradient differing from standing wave solutions of $\tilde{\theta}_{ij} =0, \pi$. When the system has periodic boundary conditions (see Fig.~\ref{Fig_FiniteVsInfinite}(d)) the pseudo-spin tilt vanishes and the lattice is characterised by a homogeneous standing wave solution (zero net-flux along $x$ and $y$ directions). The simulation in both cases uses a lattice constant of $a = 12$ $\upmu$m and is performed on a $16a \times 16a$ grid but we only plot the innermost $11a\times11a$ of the grid. Therefore, removing finite size effects results in a homogeneous power distribution among all condensate nodes and thus the phases-differences stay homogeneously $0$ or $\pi$.

We point out that by simulating a periodic system the coherence has dropped by a small amount. This can be understood from the fact that each pulse proliferates vortex solitons~\cite{Neshev_PRL2004} which, in a finite system, can decay out through the boundary of the system (i.e., damped boundary conditions destroy such defects). In the absence of such a decay channel, these defect states can survive much longer in a periodic system leading to a drop in coherence.
%
\section{EXCITATION PUMP POWER DEPENDENCE} \label{appendix_PulsedPowerDependence}
\begin{figure}[!t]
	\center
	\includegraphics[]{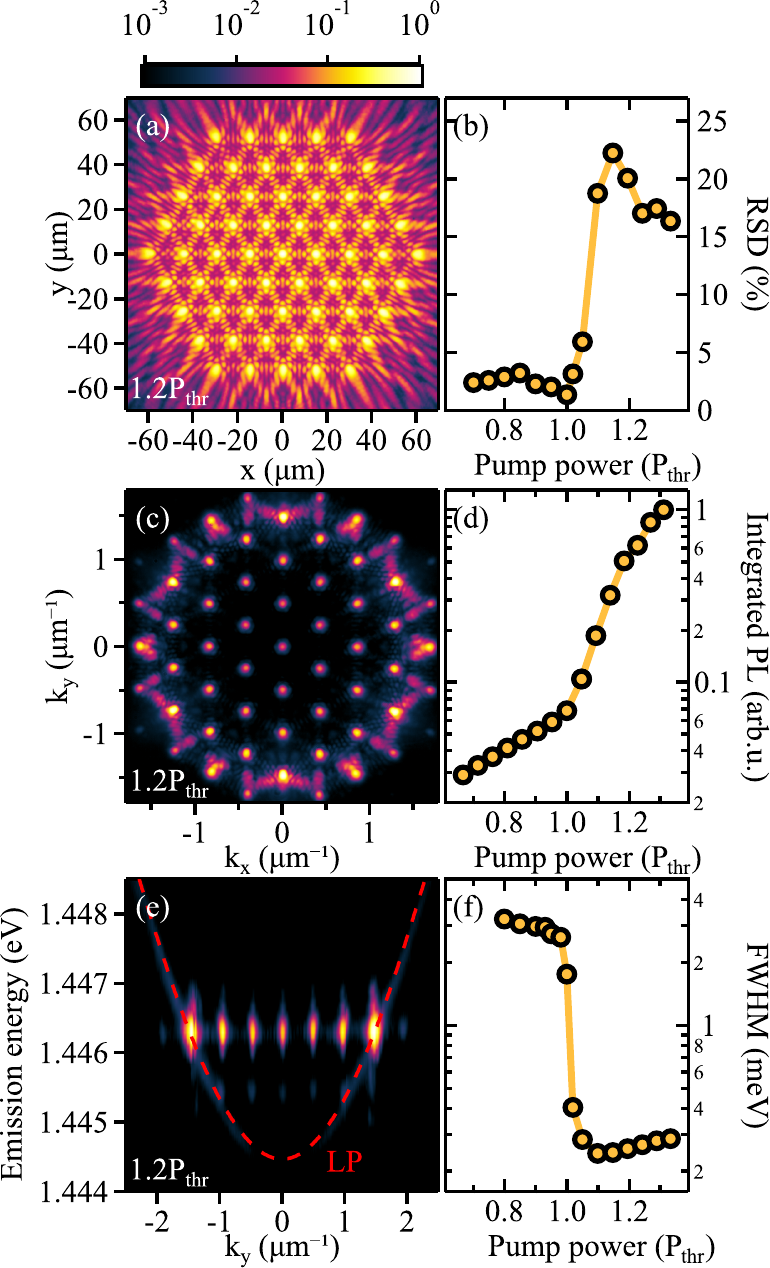}
	\caption{Pump-power dependence of a triangular lattice of polariton condensates. (a) Recorded near-field photoluminescence of 61 polariton condensates pumped at $1.2$ times condensation threshold. (b) Pump-power dependence of the relative standard deviation (RSD) of the distribution of integrated condensate emission. The system was stabilised to obtain minimum RSD ($1\%$) at condensation threshold $P_\mathrm{thr}$. (c) Recorded far-field photoluminescence for $P=1.2P_\mathrm{thr}$. (d) Pump-power dependence of the integrated emission in reciprocal space. (e) Spectrally-resolved momentum space emission along the symmetry axis $k_x = 0$ for $P=1.2P_\mathrm{thr}$. Lower polariton branch below threshold shown as red-dashed dispersion curve. (f) Pump-power dependence of the spectral full width at half maximum. }
	\label{Fig_TriangularLattice}
\end{figure}

In this section we describe the pump power dependence of the stabilised triangular lattice of 61 ballistically coupled polariton condensates with lattice constant $a=14.9\;\mathrm{\upmu m}$ shown in Figs.~\ref{Fig1} and \ref{Fig2}. For the case of pulsed excitation the pump power $P$ is a time-averaged value and a change of $P$ is equivalent to changing the peak amplitude of the sub-picosecond laser pulse. The recorded emission patterns for pump power $P=1.2P_\mathrm{thr}$ in near-field, far-field and energy resolved far-farfield along the symmetry axis $k_x=0$ are illustrated in Figs.~\ref{Fig_TriangularLattice}(a,c,e). The pump power geometry was adjusted using the described iterative feedback algorithm to stabilise the emission power of all condensates at threshold such that the spread of measured condensate emission powers shows a minimum of $1\%$ (RSD) for $P=P_\mathrm{thr}$. In Fig.~\ref{Fig_TriangularLattice}(b) we show the spread of condensate emission powers for varying total excitation pump power while keeping the relative pump power between different nodes constant. An increase of the spread above condensation threshold to about $20\%$ at $P=1.2P_\mathrm{thr}$ originates from the different number of coupled neighbouring sites between condensates in the centre and at the edge of the lattice structure leading to different additional gain from coherent coupling between condensate nodes.

In Fig.~\ref{Fig_TriangularLattice}(d) we illustrate the pump power dependence of the total integrated PL demonstrating a threshold level with non-linear increase of PL intensity which we define as condensation threshold $P_\mathrm{thr}$. An increase in pump power from condensation threshold $P_\mathrm{thr}$ to the operational point $P=1.2P_\mathrm{thr}$ leads to a 9-fold increase of total PL power.

The spectrally resolved PL in reciprocal space along the symmetry axis $k_x=0$ in Fig.~\ref{Fig_TriangularLattice}(e) demonstrates PL emission at one dominant mode with $\mathrm{FWHM} \approx 250\;\mathrm{\upmu eV}$. In agreement with the threshold-like behaviour of the integrated PL we can see a sharp narrowing of polariton linewidth at condensation threshold which is shown in Fig.~\ref{Fig_TriangularLattice}(f) .
\section{TIME-RESOLVED SYNCHRONISATION OF TWO CONDENSATES} \label{appendix_TimeResolvedEmission}
\begin{figure}[!t]
	\center
	\includegraphics[]{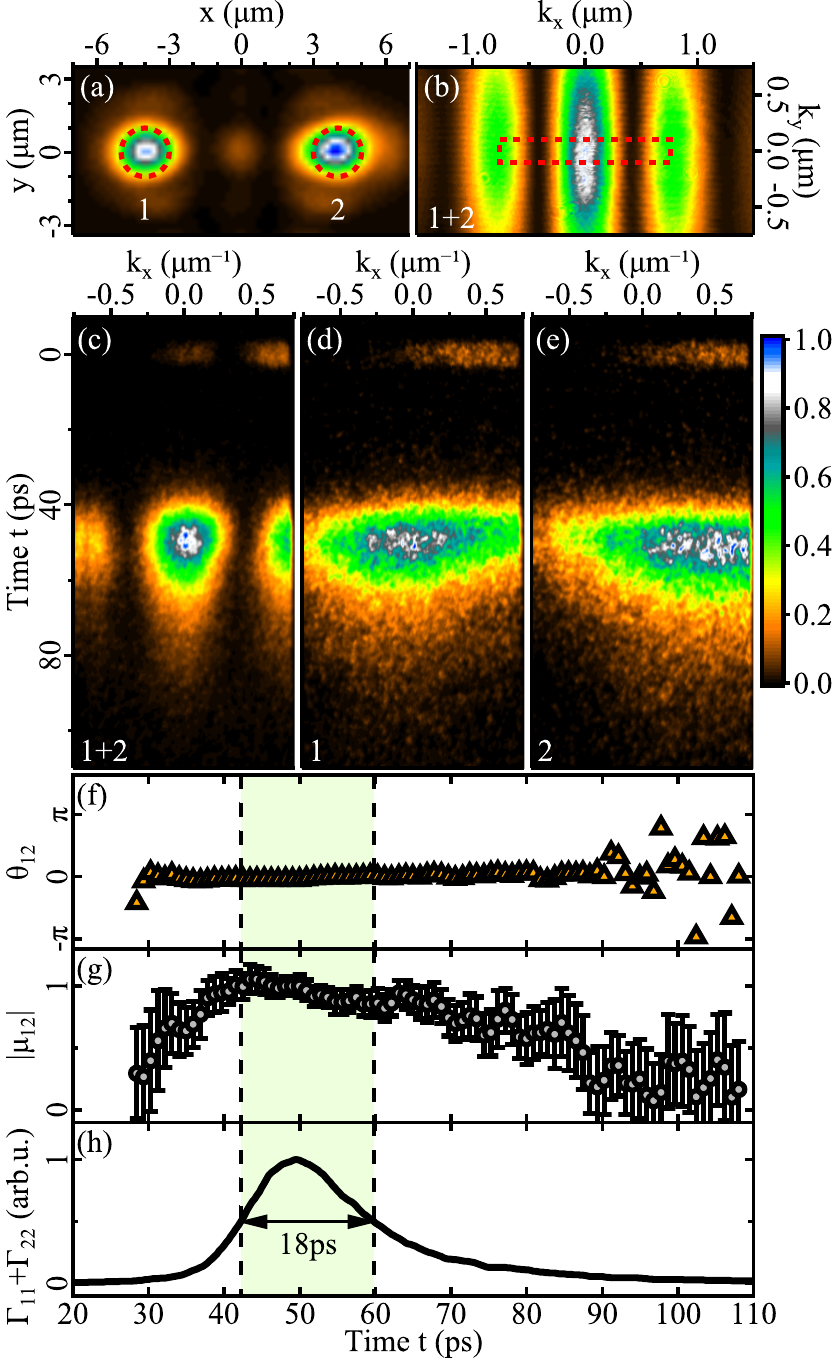}
	\caption{Synchronisation of two ballistically coupled polariton condensates. (a) Near-field photoluminescence of two condensates with separation distance $d_{12}=8\;\mathrm{\upmu m}$. (b) Recorded far-field photoluminescence when filtering the emission of the condensate centres marked with red dashed circles in (a). (c-e) Time-resolved and normalised far-field photoluminescence for both condensates interfering ($1+2$) and individually ($1,2$) recorded by projecting the photoluminescence onto the entrance slit of a streak camera (red-dashed rectangle in (b)). Time-dependencies of extracted phase difference $\theta_{12}$, the modulus of the complex degree of coherence $|\mu_{12}|$ and the integrated emission signal $\Gamma_{11}+\Gamma_{22}$ of both condensate nodes are shown in (f-h). The origin for the time axis in (f-h) is defined by the laser arrival time. The full width at half maximum of the emission signal ($\approx 18\;\mathrm{ps}$) is highlighted in light green.}
	\label{Fig_TimeResolvedDyad}
\end{figure}

The time-resolved formation of coherence in polariton condensates under non-resonant pulsed excitation has been investigated for single-~\cite{PhysRevLett.103.256402,PhysRevLett.110.137402} and two-condensate systems~\cite{Christmann_NJP2014, Ohadi_PRX2016}. In the latter case, however, a full description of the synchronisation process between two condensates in terms of their complex coherence factor has not been reported. Here, we explicitly measure the (time-resolved) complex coherence factor $\mu_{12}(t)$ of two ballistically coupled condensation centres. Pulsed excitation of two condensates with separation distance $d_{12}=8\;\mathrm{\upmu m}$  leads to the time-averaged near-field photoluminescence displayed in Fig.~\ref{Fig_TimeResolvedDyad}(a). The pump power $P\approx1.2P_\mathrm{thr}$ is the same as for the distance-dependence shown in Fig.~\ref{Fig3} in the main manuscript. One bright interference peak located in-between the two condensation centres at $x,y=0$ indicates synchronisation with vanishing phase-difference, i.e. $\tilde{\theta}_{12} = 0$. To reveal the coherence properties between the two condensates we record their far-field interference by spatially filtering the emission of both condensation centres (red-dashed circles) as shown in Fig.~\ref{Fig_TimeResolvedDyad}(b). The modulus of the integrated coherence factor from this time-averaged interference pattern is extracted as $|\tilde{\mu}_{12}| = 0.94$. By projecting the far-field pattern onto the entrance slit of a streak camera (time resolution $\Delta t \approx 2\;\mathrm{ps}$) we resolve the interference and emission of each individual condensate node in time as shown in Figs.~\ref{Fig_TimeResolvedDyad}(c-e). We extract the time-dynamics of the mutual complex degree of coherence $\mu_{12}(t)$ from the measured interference and reference signals and illustrate its phase $\theta_{12}(t)$ and modulus $|\mu_{12}(t)|$ in Figs.~\ref{Fig_TimeResolvedDyad}(f) and (g), respectively. The time-resolved occupation $\Gamma_{11}(t)+\Gamma_{22}(t)$  with $\Gamma_{ii}(t) = \langle |\psi_i(t)|^2 \rangle$ of both condensates is shown in Fig.~\ref{Fig_TimeResolvedDyad}(h) and reveals pulsed polariton emission with temporal width $\approx 18\;\mathrm{ps}$ (FWHM) and peak signal at $50\;\mathrm{ps}$ after excitation with the pump laser. We find complete synchronisation with $\theta_{12}=0$ and a maximum degree of coherence $|\mu_{12}| = 1$  appearing earlier than the peak PL signal at $t = 41\;\mathrm{ps}$ still during the condensate growth time. The subsequent temporal decay of mutual coherence is slower than the decay of total condensate population with fitted $1/e$ decay times of $80\;\mathrm{ps}$ and $13\;\mathrm{ps}$, respectively.

%
\section{COMPARISON TO CONTINUOUS WAVE EXCITATION} \label{appendix_cw}
\begin{figure}[!t]
	\center
	\includegraphics[]{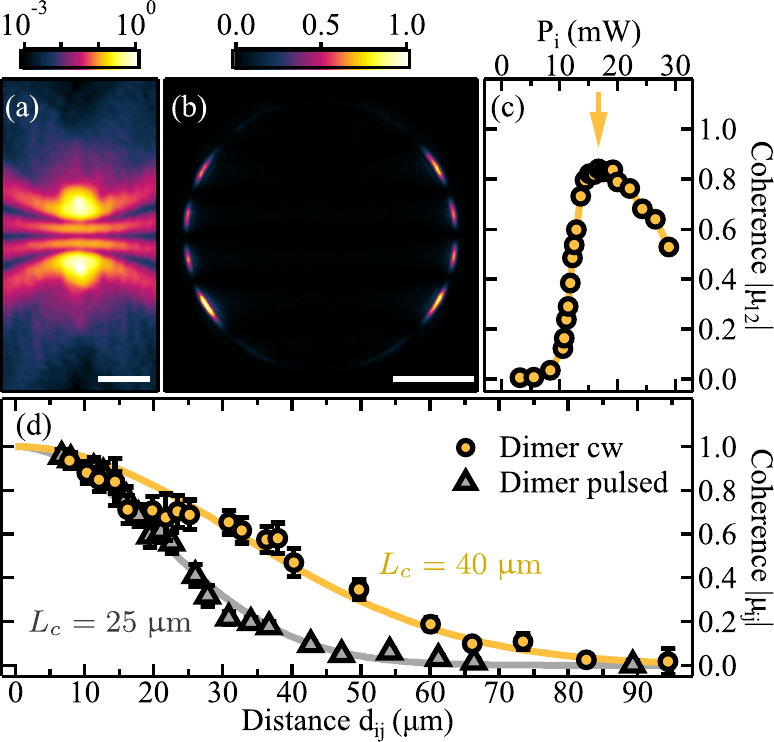}
	\caption{Ballistic coupling of two polariton condensates pumped under continuous wave (cw) excitation. Recorded (a) near-field and (b) far-field photoluminescence for two condensates separated by $d_{12}=12\;\mathrm{\upmu m}$ . (c) Extracted mutual coherence factor $|\mu_{12}|$ between the two condensates as a function of excitation pump power $P_\mathrm{i}$ per condensate. Vertical arrow marks the fixed pump power ($P\approx 1.6P_{\mathrm{thr}}$) for data shown in (a,b) and for the condensate separation distance dependence shown in (d). The data points of a dyad under pulsed excitation (grey triangles) are illustrated for comparison to the system under cw excitation (orange circles). Curves in (d) represent Gaussian fits. Scale bars in (a) and (b) correspond to $10\;\mathrm{\upmu m}$ and $1\;\mathrm{\upmu m}^{-1}$, respectively. }
	\label{Fig_CWexcitation}
\end{figure}

We compare the results of mutual coherence between two ballistically expanding polariton condensates pumped under sub-picosecond excitation shown in Fig.~\ref{Fig3} to the case of pumping using continuous wave (cw) monomode laser excitation. To prevent heating of the microcavity sample under cw excitation, the excitation laser is modulated by an acousto-optic modulator to generate square wave packets of $5\;\mathrm{\upmu s}$ duration at a frequency of $10\;\mathrm{kHz}$. In Figs.~\ref{Fig_CWexcitation}(a) and (b) we illustrate the recorded near-field and far-field photoluminescence of two condensate with separation distance $d=12\;\mathrm{\upmu m}$. The system is pumped at $P=1.6 P_{\mathrm{thr}}$ with a threshold pump power $P_{\mathrm{thr}} \approx 10\;\mathrm{mW}$ per condensate node as shown in Fig.~\ref{Fig_CWexcitation}(c). We note that the near-field and far-field excitation pattern in (a) and (b) are different to the emission patterns shown in Fig.~\ref{Fig3}(a) and (b) under pulsed excitation despite a small difference in pump spot separation distance of less than $1\;\mathrm{\upmu m}$. In both cases the emission patterns indicate synchronisation in a state with $\pi$ phase-difference between the two condensate nodes. However, the outflow wavevector (or alternatively in-plane momentum) of polaritons under pulsed excitation is larger than under cw excitation leading to a smaller interference fringe periodicity in both real- and momentum-space.

The extracted decay of the complex coherence factor $\mu_{12}$  with increasing condensate separation distance under cw excitation is illustrated in Fig.~\ref{Fig_CWexcitation}(d). In the same plot we show the dependency of the integrated complex coherence factor $\tilde{\mu}_{12}$ on the separation distance $d_{12}$ of two condensates under pulsed excitation condition (as presented in Fig.~\ref{Fig3}). While both systems demonstrate similar coherence properties for condensate separation distances $d_{12}\leq 20\;\mathrm{\upmu m}$, the stationary system's coherence $|\mu_{12}|$ is enhanced for distances larger than $20\;\mathrm{\upmu m}$ as compared to the nonstationary system under sub-picosecond pulsed excitaiton. A Gaussian fit (see Eq.~\ref{Eq.GaussianDecay}) of the decay of coherence $|\mu_{12}(d_{12})|$ with increasing condensate spacing under cw excitation yields an effective coherence length $L_C=40\;\mathrm{\upmu m}$. We argue that the build-up of coherence $|\mu_{12}(t)|$ between the two ballistically coupled condensates under pulsed excitation (as described in Appendix~\ref{appendix_TimeResolvedEmission}) is fast enough for small distances $d_{12}\leq 20\;\mathrm{\upmu m}$ as to synchronise and reach the same coherence $|\mu_{12}|$ as in steady-state operation. For larger distances, however, the increased time-of-flight of particles travelling in-between the two condensate nodes~\cite{topfer2020time} becomes noticeable and reduces the coherence factor $|\tilde{\mu}_{12}|$ of the nonstationary system with finite life-time.
%
\section{BOGOLIUBOV ANALYSIS}\label{appendix_bog}
Let us consider the continuous-wave excitation regime where $n_c(\mathbf{r}) = P(\mathbf{r})$ is taken time-independent. We will also consider the ideal case of no stochasticity by setting $dW = 0$. Assuming that the reservoir $n_x$ follows the dynamics of the condensate we can perform an adiabatic elimination of Eq.~\eqref{eq.ResA} and keeping terms to the first order in $R$ in Eq.~\eqref{eq.GP} we have,
\begin{equation}\label{eq.GP2}
i \frac{\partial \Psi}{\partial t} = \left[ -\frac{\hbar \nabla^2}{2m} + gP(\mathbf{r}) + \alpha |\Psi|^2 + \frac{iP}{2}\left(1 - \frac{|\Psi|^2}{n_s} \right) - \frac{i \gamma}{2} \right] \Psi,
\end{equation}
where $n_s = \Gamma_R/R$. Performing the standard Bogoliubov treatment where the condensate wavefunction $\Psi$ is expanded around a fixed point solution of Eq.~\eqref{eq.GP2} $\Psi_0$ with energy $\hbar \nu$ we write,
\begin{equation} \label{eq.Psi}
\Psi = \Psi_0(\mathbf{r}) e^{- i \nu t} \left[ 1 + \sum_\mathbf{q} \left(u_{n,\mathbf{q}} e^{i\mathbf{q}\cdot \mathbf{r} + \lambda_n t} +  v_{n,\mathbf{q}} e^{-i\mathbf{q} \cdot \mathbf{r} + \lambda_n^*t } \right) \right]
\end{equation}
We wish to scrutinise the dispersion of elementary excitations (Lyapunov exponents) with complex energies $\lambda_n$. Substitution of Eq.~\eqref{eq.Psi} into~\eqref{eq.GP2} and keeping only terms linear in $u_{n,\mathbf{q}}$ and $v_{n,\mathbf{q}}$ we obtain a linearised set of equations of motion for the disturbances. We assume that the potential $V$ and condensate $\Psi_0$ are infinite and periodic such that $P(\mathbf{r}) = P(\mathbf{r}+\mathbf{a})$ and $|\Psi_0(\mathbf{r})|^2 = |\Psi_0(\mathbf{r} + \mathbf{a})|^2$, where $\mathbf{a} = n_1 \mathbf{a}_1 + n_2 \mathbf{a}_2$ is the translational symmetry vector defined in the bases of primitive lattice vectors $\mathbf{a}_{1,2}$ for some integers $n_{1,2}$. We can then apply Bloch's theorem where we write the disturbances wavefunction in the factorised form of crystal momentum $\mathbf{q} = (q_x,q_y)$ and Bloch states in the $n$th band $u_{n,\mathbf{q}}(\mathbf{r})= u_{n,\mathbf{q}}(\mathbf{r}+\mathbf{a})$ and $v_{n,\mathbf{q}}(\mathbf{r})=v_{n,\mathbf{q}}(\mathbf{r}+\mathbf{a})$.
\begin{equation}
\mathbf{B}_{n,\mathbf{q}}(\mathbf{r}) = \begin{pmatrix} u_{n,\mathbf{q}}(\mathbf{r}) \\ v_{n,\mathbf{q}}(\mathbf{r}) \end{pmatrix}.
\end{equation}
By Fourier transforming periodic terms into the basis of reciprocal lattice vectors we can easily solve the energies $\lambda_n$ belonging to $u_{n,\mathbf{q}}$ and $v_{n,\mathbf{q}}$,
\begin{equation}
\mathcal{L}(\mathbf{q},\Psi_0) \mathbf{B}_{n,\mathbf{q}} = \lambda_n(\mathbf{q}) \mathbf{B}_{n,\mathbf{q}},
\end{equation}
where $\mathcal{L}$ is our Bogoliubov (Lyapunov) matrix in the crystal momentum representation,
\begin{widetext}
\begin{align}\label{eq.bloch_eig} \notag
\mathcal{L}(\mathbf{q},\Psi_0)  &= \bigg\{ \left[ \frac{\hbar }{2m}\left[ \left(q_x - i \frac{\partial}{\partial x}\right)^2 + \left(q_y - i \frac{\partial}{\partial y}\right)^2 \right] + gP(\mathbf{r}) -  \nu  + 2 \alpha |\Psi_0|^2\right] \hat{\sigma}_3 \\
& \frac{i}{2}(P - \gamma ) \hat{\sigma}_0   + \alpha \begin{pmatrix} 0 & \Psi_0^2 \\ -(\Psi_0^*)^2 & 0  \end{pmatrix} - \frac{i P}{2 n_s} \left[ 2 |\Psi|^2 \hat{\sigma}_0 +  \begin{pmatrix} 0 & \Psi_0^2 \\ (\Psi_0^*)^2 & 0  \end{pmatrix} \right] \bigg\} .
\end{align}
\end{widetext}
The solution $\Psi_0$ satisfying Eq.~\eqref{eq.GP2} can be obtained numerically using periodic boundary conditions. Plugging the obtained solution $\Psi_0$, which corresponds to observed condensate patterns in experiment, into the eigenvalue problem for $\mathbf{B}_{n,\mathbf{q}}$ we can finally diagonalise our system.
\section{FAR-FIELD DIFFRACTION FOR NARROW-BANDWIDTH PARTIALLY COHERENT LIGHT} \label{appendix_farfield_coherence}
Let us assume the narrow bandwith optical field $\Psi(\vect{r},t) = \psi(\vect{r,t}) \exp{(-i 2\pi c t /\bar{\lambda})}$ with mean wavelength $\bar{\lambda}$ is being truncated by a thin aperture with transmittance function $P(\vect{r})$ in the plane $\mathcal{A}$ such that the field directly after the aperture is given by the product $P(\vect{r}) \Psi(\vect{r},t)$. The signal's bandwidth $\Delta \nu $ is assumed to be much smaller than the central frequency $\bar{\nu} = c/\bar{\lambda}$ such that the complex amplitude $\psi(\vect{r},t)$ is a slowly varying envelope in time. The average intensity distribution $\tilde{I}(\vect{q})$ of the resulting far-field diffraction pattern described in coordinate basis $\vect{q}$ - realised at the back focal plane of a thin Fourier-transforming lens with focal length $f$ -   in analogy to Schell's theorem~\cite{goodman2015statistical} can be approximated as
\begin{equation}
\begin{aligned}
\label{eq.SchellsTheorem}
&\tilde{I}(\vect{q}) = \\ &
\iint\displaylimits_{\mathcal{A}} \iint\displaylimits_{\mathcal{A}} P^*(\vect{r}_1) P(\vect{r}_2)  \sqrt{\tilde{I}(\vect{r}_1) \tilde{I}(\vect{r}_2)} \tilde{\mu}(\vect{r}_1,\vect{r}_2)  \frac{\mathrm{e}^{i \frac{2\pi}{\bar{\lambda} f} \vect{q} \cdot \vect{d}_{12}}}{(\bar{\lambda} f)^2}  \mathrm{d}\vect{r}_1\mathrm{d}\vect{r}_2 ,
\end{aligned}
\end{equation}
with distance vector $\vect{d}_{12} = \vect{r}_1 - \vect{r}_2$, integrated complex coherence factor $\tilde{\mu}(\vect{r}_1,\vect{r}_2)$ and average intensity $\tilde{I}(\vect{r}_{1,2})$ of the optical field $\psi(\vect{r_{1,2},t})$ at locations $\vect{r}_{1,2}$ in the input plane $\mathcal{A}$. We further approximate a distribution of point-like holes in the aperture, i.e.
\begin{equation}
\label{eq.HoleAperture}
P(\vect{r}) = A_0 \sum_i \delta \left( \vect{r} - \vect{r}_i \right),
\end{equation}
 where $A_0$ corresponds to the finite physical size of each hole, and we transform into the basis of spatial frequencies $\vect{k}(\vect{q}) = 2\pi\vect{q}/\bar{\lambda} f$ yielding
\begin{equation}
\label{eq.SchellsDiscreteTheorem}
\tilde{I}(\vect{k}) =  \frac{A_0^2}{ (\lambda f)^2}    \sum_{i,j} \sqrt{\tilde{I}(\vect{r}_i) \tilde{I}(\vect{r}_j)} \tilde{\mu}(\vect{r}_i,\vect{r}_j) \mathrm{e}^{i  \vect{k} \cdot \vect{d}_{ij}} .
\end{equation}
Under the assumption of a fully coherent field with constant coherence factor $\tilde{\mu}(\vect{r}_i,\vect{r}_j) = 1$ the resultant far-field diffraction pattern (Eq.~\eqref{eq.SchellsDiscreteTheorem}) reduces to the well-known Fraunhofer diffraction formula,
\begin{equation}
\label{eq.DiscreteFraunhoferDiffractionFormula}
\tilde{I}(\vect{k}) = \left|  \frac{A_0}{ \lambda f}    \sum_{i}  \sqrt{\tilde{I}(\vect{r}_i)} \mathrm{e}^{i  \vect{k} \cdot \vect{r}_{i}}  \right|^2.
\end{equation}
It is apparent from Eq.~\eqref{eq.DiscreteFraunhoferDiffractionFormula} that in case of a periodic arrangement (lattice) of aperture holes at locations $\vect{r}_i$ the resultant intensity distribution $\tilde{I}(\vect{k})$ is formed by the squared modulus of the discrete Fourier-transform of the optical field sampled at locations $\vect{r}_i$.
\newpage
\bibliography{refs}

\end{document}